\documentclass[fleqn,usenatbib]{mnras}
\usepackage{newtxtext,newtxmath}
\usepackage{amsmath}
\usepackage[T1]{fontenc}
\usepackage{ae,aecompl}

\usepackage[final]{graphicx}	
\usepackage{amsmath}	
\usepackage{amssymb}	
\usepackage{color}

\title[]{Astronomical masers and Dicke's superradiance}
\author[Rajabi \& Houde]{Fereshteh Rajabi,$^{1,2}$\thanks{E-mail: f3rajabi@uwaterloo.ca}
Martin Houde$^{3}$\thanks{E-mail: mhoude2@uwo.ca}
\\
$^{1}$Perimeter Institute for Theoretical Physics, Waterloo, ON N2L 2Y5, Canada\\
$^{2}$Institute for Quantum Computing and Department of Physics and Astronomy, The University of Waterloo, 200 University Ave. West, \\Waterloo, Ontario N2L 3G1, Canada\\
$^{3}$Department of Physics and Astronomy, The University of Western Ontario, 1151 Richmond Street, London, Ontario N6A 3K7, Canada\\}

\date{}

\pubyear{2019}

\begin{document}
\label{firstpage}
\pagerange{\pageref{firstpage}--\pageref{lastpage}}
\maketitle
%
\begin{abstract}
We consider the radiation properties and processes of a gas with a population inversion using the formalism based on the Maxwell-Bloch equations. We focus on the maser action and Dicke's superradiance to establish their relationship in the overall radiation process during the temporal evolution of the system as a function of position. We show that the maser action and superradiance are not competing phenomena but are rather complementary, and define two distinct limits for the intensity of radiation. Masers characterise the quasi-steady state limit, when the population inversion density and the polarisation amplitude vary on time-scales longer than those of non-coherent processes affecting their evolution (e.g., collisions), while superradiance defines the fast transient regime taking place when these conditions are reversed. We show how a transition from a maser regime to superradiance will take place whenever a critical threshold for the column density of the population inversion is reached, at which point a strong level of coherence is established in the system and a powerful burst of radiation can ensue during the transient regime. This critical level also determines the spatial region where a transition from the unsaturated to the saturated maser regimes will take place; superradiance can thus be seen as the intermediary between the two. We also quantify the gain in radiation intensity attained during the superradiance phase relative to the two maser regimes, and show how the strong coherence level during superradiance is well suited to explain observations that reveal intense and fast radiation flares in maser-hosting regions.           
\end{abstract}


\section{Introduction}

Astronomical masers have been the subject of great interest since their first detection in 1965 \citep{Weaver1965}. By their very nature and the requirements needed for their existence they reveal invaluable information about their host environments. Several spectral transitions are available for such studies, as the maser action is routinely detected in a wide range of molecular species and lines (e.g., 6.7~GHz CH$_3$OH, 1612/1665/1667/1712~MHz OH, 22~GHz H$_2$O, etc.) and a variety of types of sources (e.g., star-forming regions, the circumstellar envelopes of evolved stars and Active Galactic Nuclei; see \citealt{Gray2012}). 

The underlying physical mechanism behind the formation of masers is, as its name implies, the amplification of radiation through stimulated emission. In a nutshell, in a masing region, photons propagating through a gas of excited molecules will stimulate the emission of photons of similar phase, frequency, and polarisation along velocity coherent paths, resulting in strongly beamed emission. The corresponding radiation is detected over regions spanning only a few astronomical units in size on the sky, while exhibiting brightness temperatures as high as $10^{12}-10^{13}\,\mathrm{K}$ \citep{Gray2012}. Moreover, the monitoring of maser-hosting regions reveals flux variations lasting over a wide domain of time-scales, ranging from weeks \citep{Goedhart2003, Goedhart2004,Szymczak2018a} to decades \citep{Bloemhof1992, Brand2018}. These events are associated with variations in the pumping source responsible for the population of excited molecules (i.e., the level of inversion in the gas), which, in principle, can be traced to changes in the physical conditions of the local environments. However, there are some cases where light curves exhibiting unusual fast rises in intensity or complex light profiles are observed. To explain these observations, for instance flaring in some velocity components of S255IR-NIRS 3 where an increase of more than a factor of a 1000 was recorded over a time span of less than 100 days \citep{Rajabi2019, Szymczak2018b}, the volume of the emitting gas or the pump rate of the source should increase by a large factor over a similar time span according to maser theory alone \citep{Szymczak2018c, Szymczak2018b}. Explaining such fast changes in a source can be challenging, necessitating more in-depth studies. 

Another energetic phenomenon in which a group of molecules in an excite state interact with a common radiation field and release their stored energy as a directional intense beam is superradiance, first introduced by R. H. Dicke in 1954 \citep{Dicke1954}. Over the last several decades an abundant literature has developed on the theory and experimental realisations of superradiance in different gaseous or solid state laboratory systems (see \citealt{Gross1982, Benedict1996, Cong2016} for reviews). Recently, \citealt{Rajabi2016B,Rajabi2017,Rajabi2019} provided some evidence for superradiance in astronomical media, specifically in regions known for harbouring masers. 

The fundamental difference between superradiance and the maser action arises from the mechanism through which photons are generated. In superradiance, a sample of $N$ excited molecules sees them become entangled through their interaction with a common radiation field, which acts as a mediator. The entangled sample then behaves as a single quantum mechanical system, damping away its stored energy at once through an enhanced spontaneous emission process. In the ideal case, the corresponding rate of the enhanced spontaneous emission in Dicke's superradiance is $N\Gamma$, where $\Gamma$ is the spontaneous emission rate for a single molecule. This is in contrast to the masing action, where there is no modified spontaneous emission rates, and the output signal is simply the product of a series of localised stimulated emission events.

With this paper our main goal is to clarify the relationship between the masing action and superradiance within the context of an inverted radiating gas, i.e., a radiating gas that sustains a population inversion. We seek to answer questions pertaining to their roles and importance for different phases in the radiation process.  For example, are the masing action and superradiance competing phenomena, or do they take place independently in different regimes? When, within the context of astronomical media, and under what conditions can entangled quantum mechanical states be established in an inverted gas leading to superradiance? Anticipating some answers for these questions, we can also ask how and when will a system transition between the maser and superradiance regimes? 

To address these questions, we have outlined the paper as follows. In Section \ref{sec:theory}, we provide a simple theoretical framework where we show that the masing action and superradiance can both be described with the so-called Maxwell-Bloch equations, and emerge as two distinct limits within this formalism. In Section \ref{sec:examples}, we analyse two scenarios (constant and pulsed inversion pumping) through numerical computations to better establish the relationship between masers and superradiance (Sections \ref{subsec:step} and \ref{subsec:pulse}), and apply our formalism to actual data. Finally, in Section \ref{sec:conclusion} we summarise our results, while two appendices, providing some detailed calculations, will be found at the end.

\section{Theoretical framework}\label{sec:theory}

The interaction between a radiation field and a group of molecules can be described using the Maxwell-Bloch equations

\begin{align}
     & \frac{\partial\hat{n}^\prime_{v}}{\partial\tau} = \frac{i}{\hbar}\left(\hat{P}_{v}^+\hat{E}^+-\hat{E}^-\hat{P}_{v}^-\right)-\frac{\hat{n}^\prime_v}{T_1}+ \hat{\Lambda}_\mathrm{n} \label{eq:dN/dt} \\
     & \frac{\partial\hat{P}_{v}^+}{\partial\tau} = \frac{2id^2}{\hbar}\hat{E}^-\hat{n}^\prime_v-\left(\frac{1}{T_2} - ikv\right)\hat{P}_{v}^+ + \hat{\Lambda}_{\mathrm{p}} \label{eq:dP/dt} \\
     & \frac{\partial\hat{E}^+}{\partial z} = \frac{i\omega_0}{2\epsilon_0c} \int dv F\left(v\right) \hat{P}_{v}^-\label{eq:dE/dt},
\end{align}
where $\hat{n}^\prime_v$ is (half of) the inverted population density, while $\hat{P}_{v}^+$ and $\hat{E}^+$ are the amplitudes of the molecular polarisation and the electric field component of the interacting field, respectively. Here, the superscript ``$+$'' stands for the polarization associated with the molecular transition from a lower level to an upper level and the positive frequency component of the electric field. It should be noted that in this framework the molecules are approximated as two-level systems, and the subscript ``$v$'' is used to identify the velocities at which different groups of excited molecules move in the sample. In the derivation of equations (\ref{eq:dN/dt})-(\ref{eq:dE/dt}), the slowly varying envelope approximation (SVEA) was used \citep{Arecchi1970, MacGillivray1976, Gross1982,Benedict1996,Rajabi2016B}, where the polarisation and electric field were modelled with 

\begin{align}
& \mathbf{\hat{P}}_v^\pm\left(z ,\tau, v\right) = \hat{P}_v^\pm\left(z ,\tau, v\right) e^{\pm i\omega_0\tau}\boldsymbol{\epsilon}_\mathrm{d}\label{eq:penvelope}\\
& \mathbf{\hat{E}^\pm}\left(z ,\tau\right) = \hat{E}^\pm\left(z ,\tau\right) e^{\mp i\omega_0\tau}\boldsymbol{\epsilon}_\mathrm{d}\label{eq:Eenvelope},
\end{align}
with $\boldsymbol{\epsilon}_\mathrm{d} = \mathbf{d}/d$ the unit polarisation vector associated with the molecular transition at a frequency $\omega_0 = c k$ and a transition dipole moment $d=\left| \mathbf{d}\right|$. The molecules in the system are excited due to a pump operating at a rate characterized by $\hat{\Lambda}_\mathrm{n}$, while random spontaneous emission events throughout the sample act as a polarisation source $\hat{\Lambda}_{\mathrm{p}}$ at which rate $\hat{P}_{v}^+$ is produced. An external field can also induce polarisation in the sample in which case its effect could also be incorporated into $\hat{\Lambda}_{\mathrm{p}}$. Losses in excitation or polarisation in the molecular sample via non-coherent processes (e.g., collisions) are modelled, respectively, with the time-scales $T_1$ and $T_2$. The Doppler shifts resulting from molecular motions are accounted for by the term $ikv$ in equation (\ref{eq:dP/dt}) for the temporal rate of change in polarisation, while in equation (\ref{eq:dE/dt}) for the electric field their effect is contained in the integration of the polarisation over the corresponding velocity distribution profile $F(v)$. 

Equations (\ref{eq:dN/dt})-(\ref{eq:dE/dt}) describe the evolution of a so-called one-dimensional system, which is well adapted to the long cylindrical representation we will be using for a radiating gas. The lone spatial dependency is therefore in the direction of propagation (i.e., along the $z$-axis), while temporal evolution is tracked using the retarded time $\tau=t-z/c$, with $c$ the speed of light. It should be noted that although we derived equations (\ref{eq:dN/dt})-(\ref{eq:dE/dt}) within a fully quantum mechanical formalism using the Heisenberg picture, other approaches can be used (e.g., through the density matrix formalism) and they very closely resemble similar classical Maxwell-Bloch equations derived for interacting matter-field systems (see \citealt{MacGillivray1976, Gross1982, Benedict1996} for details). Since a classical formalism is more practical (e.g., the exact order of the variables is not followed) and is usually adopted for numerical computations purposes, we omit the operator sign (i.e., the ``caret'' notation $\hat{X}$ for a quantity $X$) from now on. 

Most importantly for our analysis, we are interested in two distinct limits in the radiation process that will be shown to emerge from equations (\ref{eq:dN/dt})-(\ref{eq:dE/dt}) depending on the variations of $P_{v}^+$ and $n^\prime_v$ as a function of the retarded time $\tau$. These limits describe two different radiation regimes: one describing the masing action and another pertaining to superradiance. More precisely, when the time-scales for variations in the polarisation amplitude and the population inversion density are longer than $T_1$ and $T_2$ in equations (\ref{eq:dN/dt})-(\ref{eq:dP/dt}), the system is in the so-called quasi-steady state limit and radiation occurs through the maser emission process. Conversely, superradiance will only set in for the so-called fast transient limit when the changes in $P_{v}^+$ and $n^\prime_v$ happen over time-scales short relative to $T_1$ and $T_2$. We now investigate these modes of radiation in the following two sections, starting with the quasi-steady state regime. 

To simplify our discussion, while still retaining the essential physics of the problem, we will consider our system to be at resonance and assume no contribution from polarisation sources. That is, we set $\Lambda_\mathrm{p}=0$, while we neglect inhomogeneous (Doppler) broadening by setting $kv=0$ and $F\left(v\right)=\delta\left(v\right)$ in equations (\ref{eq:dP/dt})-(\ref{eq:dE/dt}). We can therefore substitute $n^\prime_v\rightarrow n^\prime$ and $P_{v}^{\pm}\rightarrow P^{\pm}$ without any risk of confusion. Otherwise, our exposition will closely follow those found in the quantum optics literature on superradiance (e.g., \citealt{Feld1980,Gross1982,Benedict1996}) and our earlier papers on the subject \citep{Rajabi2016A,Rajabi2016B,Rajabi2017,Houde2018a,Houde2018c,Houde2019,Rajabi2019}.

\subsection{The quasi-steady state limit - the maser action}\label{subsec:quasisteady}

As previously mentioned, and following the treatment of \citet{Feld1980}, the quasi-steady state limit is applicable when temporal variations in polarisation amplitude and population inversion density happen on long time-scales compared to those characterising the relaxation and dephasing processes, i.e.,
\begin{equation}
\frac{\partial n^\prime}{\partial\tau} \ll \frac{n^\prime}{T_1} \quad\mathrm{and}\quad \frac{\partial P^+}{\partial\tau} \ll \frac{P^+}{T_2}.\label{eq:steady-state} 
\end{equation}
We thus neglect the time derivatives on the left-hand side of equations (\ref{eq:dN/dt}) and (\ref{eq:dP/dt}) to find 
\begin{align}
& n^\prime = \frac{n^\prime_0}{1 + \left|\frac{2dE^+}{\hbar}\right|^2 T_1 T_2} \label{eq:Nsteady}\\
& P^+ = \left(\frac{2id^2T_2}{\hbar}\right)E^- n^\prime,\label{eq:Psteady}
\end{align}
where $n^\prime_0 = \Lambda_\mathrm{n} T_1$ is (half) the inverted population density in the absence of radiation (i.e., when $E^\pm = 0$). We assume for the moment that $\Lambda_\mathrm{n}$ is constant over time. Equation (\ref{eq:Nsteady}) can be rewritten as
\begin{equation}
n^\prime = \frac{n^\prime_0}{1 + I/I_{\mathrm{sat}}}, \label{eq:Nsteady2}
\end{equation}
where $I = c\epsilon_0|E^+|^2/2$ is the intensity of output radiation and
\begin{equation}
I_{\mathrm{sat}} = \frac{c\epsilon_0 \hbar^2}{8d^2 T_1T_2} \label{eq:Is} 
\end{equation}
is the saturation intensity. 

Inserting equation (\ref{eq:Psteady}) into equation (\ref{eq:dE/dt}) (at resonance) and applying the definition for the intensity, one can write
\begin{equation}
\frac{dI}{dz} = \frac{2\omega_0d^2 T_2 n^\prime}{c\epsilon_0\hbar}I. \label{eq:IntensityMaser}
\end{equation}
When accounting for the dependency on $I$ of $n^\prime$ in equation (\ref{eq:Nsteady}), a solution for equation (\ref{eq:IntensityMaser}) is expressible using Lambert's W function \citep{Gray2012}. It will, however, be sufficient for us to explore two limiting regimes in the intensity. We thus define a new parameter
\begin{equation}
\alpha = \frac{\omega_0d^2T_2 n }{c\epsilon_0\hbar},\label{eq:alpha} 
\end{equation}
with $\alpha>0$ the gain coefficient for the intensity as the radiation propagates through the medium, and $n=2 n^\prime$ the full inversion level. We thus transform equation (\ref{eq:IntensityMaser}) to
\begin{equation}
\frac{dI}{dz} = \alpha I. \label{eq:IntensityMaser2}
\end{equation}

In the weak field limit, i.e., when $I\ll I_{\mathrm{sat}}$, $\alpha$ is a constant and equation (\ref{eq:IntensityMaser2}) admits a solution in the form of
\begin{equation}
I(z)  = I_0 e^{\alpha z},\label{eq:IntensityMaserWeak}
\end{equation}
where $I_0$ is the (background) intensity at $z=0$. The exponential growth in intensity with $z$ appearing in equation (\ref{eq:IntensityMaserWeak}) is a characteristic of unsaturated masers.  

At the opposite end in the strong field limit, when $I\gg I_{\mathrm{sat}}$, $\alpha$ becomes inversely proportional to $I$ through its dependency on the population inversion level (see equation (\ref{eq:Nsteady})). As a result the spatial variation in intensity $dI/dz$ becomes independent of $I$ and admits the following linear solution
\begin{equation}
    I(z) \simeq \frac{\hbar\omega_0 n_0}{8 T_1}z,\label{eq:IntensityMaserStrong}
\end{equation}
where we set $n_0=2 n^\prime_0$ for the full inverted population level in the absence of radiation. This linear growth is a characteristic of a saturated maser, where the gain scales linearly with $n_0$ and the intensity is a linear function of the position $z$ in the medium.

Although the stimulated emission process is not apparent in the Maxwell-Bloch equations (i.e., equations (\ref{eq:dN/dt})-(\ref{eq:dE/dt})), we can still surmise from the previous analysis and the corresponding functionalities that their quasi-steady state solutions in the weak and strong field limits do pertain to the masing action phenomenon. That is, the natural occurrence in the formalism of a saturation intensity $I_\mathrm{sat}$ and consequently that of an exponential growth (weak intensity), as well as a linear growth (strong intensity) is in perfect correspondence with results one obtains using the maser rate equations, where stimulated emission plays a central role \citep{Elitzur1992, Gray2012}. 

\subsection{The fast transient limit - superradiance}\label{subsec:transient}

Rapid variations of polarisation and population inversion density on time-scales shorter than $T_1$ and $T_2$ benchmark the fast transient limit of the Maxwell-Bloch equations. More precisely, we now consider situations where

\begin{equation}
\frac{\partial n^\prime}{\partial\tau} \gg \frac{n^\prime}{T_1} \quad\mathrm{and} \quad \frac{\partial P^+}{\partial\tau} \gg \frac{P^+}{T_2}. \label{eq:transientlimit}
\end{equation}
Such rapid variations of $P^+$ and $n^\prime$ in an inverted system allow the realisation of a coherent regime in the radiation process that otherwise would have been prohibited by non-coherent relaxation/dephasing effects. Importantly for what will follow, superradiance as a coherent phenomenon can proceed when
\begin{equation}
T_{\mathrm{R}} \ll T_1, T_2,
\end{equation}
with $T_{\mathrm{R}}$ its characteristic time-scale soon to be defined. 

When conditions (\ref{eq:transientlimit}) are met in a system, the Maxwell-Bloch equations (still at resonance) are simplified to  
\begin{align}
& \frac{\partial n^\prime}{\partial\tau} = \frac{i}{\hbar}\left(P^+E^+-E^-P^-\right)\label{eq:dN/dt-Ideal} \\   
& \frac{\partial P^+}{\partial\tau} = \frac{2id^2}{\hbar}E^-n^\prime  \label{eq:dP/dt-ideal} \\
& \frac{\partial E^+}{\partial z} = \frac{i\omega_0}{2\epsilon_0c} P^-\label{eq:dE/dt-ideal},
\end{align}
corresponding to the so-called ideal case when non-coherent processes are not present. We also set the population inversion and polarisation pumps to zero while we instead impose initial conditions for $n^\prime$ and $P^+$ at $\tau=0$ (this is equivalent to assuming an instantaneous inversion at $\tau = 0$ at all positions $z$). It should also be noted that the notion of an instantaneous inversion is an idealisation which, although approachable in laboratory experiments, is unlikely to be achieved in astronomical media. We use it here, and again in Section \ref{subsec:step}, to simplify our discussion in order to more clearly bring out the essential physics of superradiance and its complementary relationship with the maser action. The treatment of a more realistic case will follow in Section \ref{subsec:pulse}.

Under these conditions equations (\ref{eq:dN/dt-Ideal})-(\ref{eq:dP/dt-ideal}) admit the following solution
\begin{align}
& n^\prime = n^\prime_0 \cos{\left(\theta\right)}\label{eq:Ncosine}\\
& P^+ = n^\prime_0 d \sin{\left(\theta\right)}\label{eq:Nsine},
\end{align}
where, as before, $n^\prime_0 = n^\prime\left(z,\tau = 0\right)$ and the so-called Bloch angle $\theta$ obeys (e.g., from equation (\ref{eq:dP/dt-ideal})) 
\begin{equation}
\frac{d\theta}{d\tau} = \frac{2id}{\hbar}E^-. \label{eq:theta1}
\end{equation}

Upon further applying the change of variable
\begin{equation}
    q = 2\sqrt{\frac{z\tau}{L T_\mathrm{R}}}\label{eq:q}
\end{equation}
with $L$ the length of the (cylindrical) system and 
\begin{equation}
    T_\mathrm{R} = \tau_\mathrm{sp}\frac{8\pi}{3\lambda^2 n_0L}\label{eq:TR}
\end{equation}
equations (\ref{eq:dE/dt-ideal}) and (\ref{eq:theta1}) can be reduced to the so-called sine-Gordon equation  
\begin{equation}
\frac{d^2 \theta}{dq^2} + \frac{1}{q}\frac{d\theta}{dq} = \sin(\theta),\label{eq:theta2ndOrder}
\end{equation}
where $\theta = \theta(q)$. In equation (\ref{eq:TR}) $\tau_\mathrm{sp}$ is the spontaneous emission time-scale of the molecular transition under consideration (i.e., the inverse of the Einstein spontaneous emission coefficient) and $\lambda$ the wavelength of the radiation field. We also note that the characteristic time-scale of superradiance is related to the gain of an unsaturated maser through
\begin{equation}
    \alpha = \frac{2 T_2}{L T_\mathrm{R}}.\label{eq:alpha-TR}
\end{equation}

For a system starting with an inverted population density $n_0=2 n^\prime_0$ the radiation field will be initiated as some molecules start to spontaneously emit photons, building up a small polarisation in the process. It can be shown that the initial Bloch angle $\theta_0$ resulting from these fluctuations will be a function of the number of molecules $N$ through $\theta_0\simeq2/\sqrt{N}\ll 1$ \citep{Gross1982}.

\begin{figure}
    \centering
    \includegraphics[width=\columnwidth]{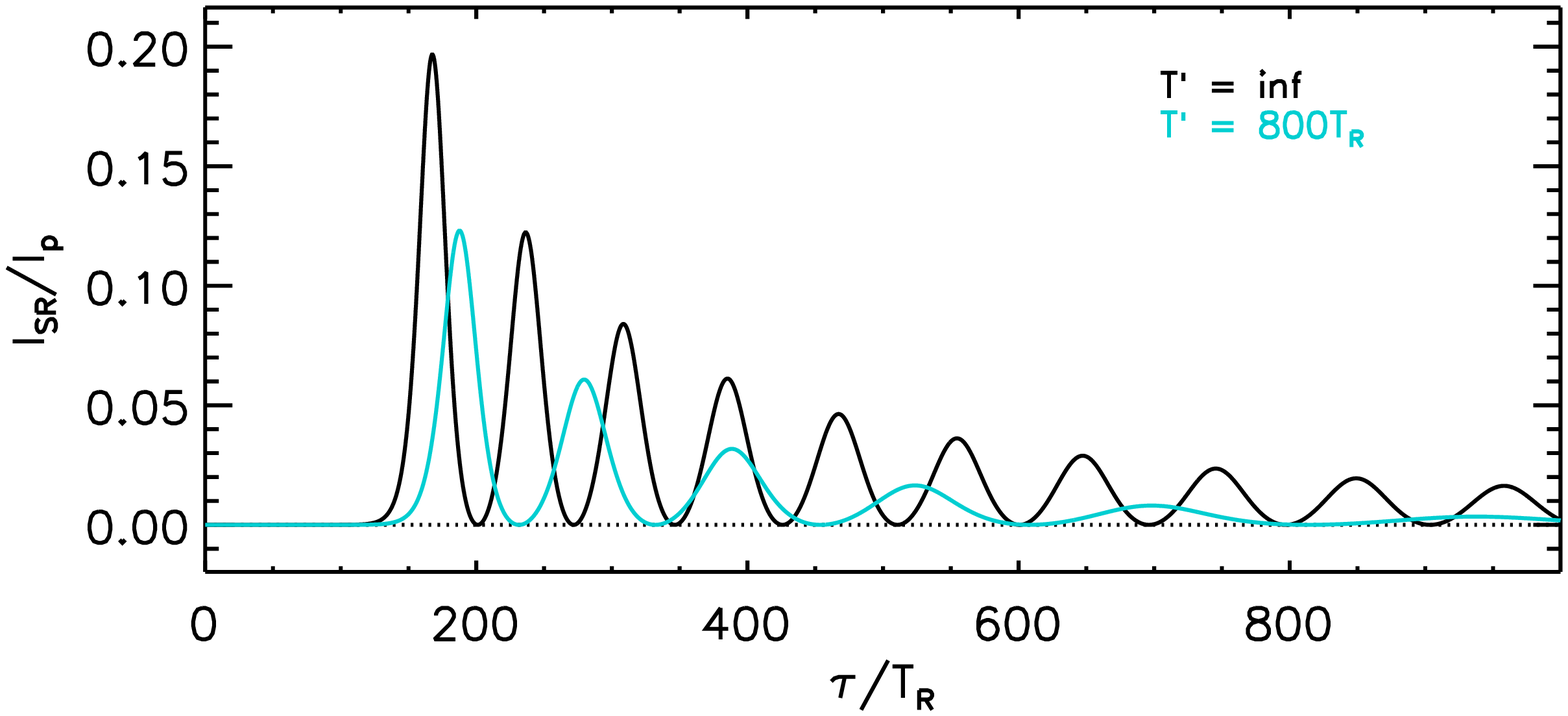}
    \caption{Superradiance intensity curves at the 6.7~GHz methanol transition for the ideal case (dark/black curve), i.e., with $T^\prime=\infty$, and for a system with damping due to non-coherent relaxation and dephasing processes (light/cyan curve; $T^\prime=800\,T_\mathrm{R}$). The intensity is scaled to the (approximate) theoretical expectation for the peak intensity (equation (\ref{eq:PeakIntensitySR})) and the retarded time $\tau$ to the characteristic superradiance time-scale $T_\mathrm{R}$ (equation (\ref{eq:TR})). In both cases we modelled the system with a cylinder of length $L=2\times 10^{15}\,\mathrm{cm}$ and radius $w=5.4\times 10^7\,\mathrm{cm}$ (corresponding to a Fresnel number of unity). The initial population inversion density corresponds to a total inversion of approximately $0.1\,\mathrm{cm^{-3}}$ for a molecular population spanning 1~km~s$^{-1}$ in velocity. With these parameters the characteristic time-scale $T_\mathrm{R}=4\times 10^4\,\mathrm{s}$.}
    \label{fig:I/Ip}
\end{figure}

Equation (\ref{eq:theta2ndOrder}) can be solved numerically for $\theta_0$ and $d\theta/d q=0$ at $q=0$ (this is equivalent to setting $E^+\left(z=0,\tau\right)=E^+\left(z,\tau=0\right)=0$). An example for the intensity of radiation at the end-fire (i.e., at $z=L$) of a cylindrical assembly of methanol molecules radiating at the 6.7~GHz spectral transition ($\lambda=4.5\,\mathrm{cm}$ and $\tau_\mathrm{sp}=6.4\times10^8\,\mathrm{s}$) is shown in Figure \ref{fig:I/Ip} (dark/black curve). The length of the cylinder was set to $L=2\times 10^{15}\,\mathrm{cm}$ with a radius $w=5.4\times 10^7\,\mathrm{cm}$ (for a Fresnel number of unity, i.e., $A=\lambda L$ with $A=\pi w^2$ the cross-sectional area of the cylinder) and the initial population inversion density to $n_0=3.3\times 10^{-12}\,\mathrm{cm^{-3}}$, corresponding to a total inversion of approximately $0.1\,\mathrm{cm^{-3}}$ for a molecular population spanning 1~km~s$^{-1}$ in velocity \citep{Rajabi2019}. Under these conditions the characteristic time-scale of superradiance is $T_\mathrm{R}=4\times 10^4\,\mathrm{s}$ from equation (\ref{eq:TR}) and the number of molecules involved in the radiation process is $N=6\times 10^{19}$. In the figure the intensity is normalised to the expected peak intensity $I_\mathrm{p}$ (see equation (\ref{eq:PeakIntensitySR}) below), while the retarded time $\tau$ is scaled to $T_\mathrm{R}$. The intensity curve is characterised by the appearance of a series of intensity bursts, a phenomenon known as the ``ringing effect,'' following a time delay during which the intensity is negligible. Equation (\ref{eq:theta2ndOrder}) can be solved analytically for small $\theta$ \citep{Benedict1996}, from which it can be shown that the first burst has a peak intensity \citep{MacGillivray1976,Feld1980}
\begin{equation}
I_{\mathrm{p}} \approx \frac{4N\hbar\omega_0/\left(AT_{\mathrm{R}}\right)}{\left|\ln\left(\theta_0/2\pi\right)\right|^2}
\propto N^2 \label{eq:PeakIntensitySR}
\end{equation}
while the time delay is 
\begin{equation}
\tau_{\mathrm{D}} \approx \frac{T_{\mathrm{R}}}{4}\left|\ln\left(\frac{\theta_0}{2\pi}\right)\right|^2 \propto N^{-1}. \label{eq:tau_D}
\end{equation}
The dependency on the number of molecules $N$ in equations (\ref{eq:PeakIntensitySR})-(\ref{eq:tau_D}) follows from re-writing the superradiance time-scale (equation (\ref{eq:TR})) as 
\begin{equation}
    T_\mathrm{R} = \tau_\mathrm{sp}\frac{8\pi}{3\phi_\mathrm{D} N}\label{eq:TR-2}
\end{equation}
with $\phi_\mathrm{D}=\lambda^2/A$ the diffraction angle at the end-fire ($z = L$) of the cylinder. The intensity curve for the ideal superradiance system (i.e., when the effects of $T_1$ and $T_2$ are negligible) agrees reasonably well with the approximate equations (\ref{eq:PeakIntensitySR}) and (\ref{eq:tau_D}) for the peak intensity (i.e., within a factor of a few) and the time delay before the appearance of a susbtantial superradiance signal, as we calculate $\tau_\mathrm{D}\approx143\,T_\mathrm{R}$ from the corresponding equation.  

Although the entangled quantum mechanical states used by \citet{Dicke1954} (i.e., the Dicke states) to derive the basic properties of superradiance and coherent radiation from a gas of excited molecules are not explicit in the Maxwell-Bloch equations (i.e., equations (\ref{eq:dN/dt-Ideal})-(\ref{eq:dE/dt-ideal})), the characteristics of superradiance are contained within them. That is, the features previously discussed, i.e., the scaling of $I_{\mathrm{p}}$ with the square of the number of inverted molecules $N$, and the existence of a characteristic time-scale $T_\mathrm{R}$ as well as a time delay $\tau_\mathrm{D}$ both inversely proportional to $N$ are all signatures of superradiance. 

Finally, we note that the characteristic time-scale of superradiance $T_\mathrm{R}$, and to a large extent the overall response of the system, is dependent on the column density of the inverted population $n_0 L$. In other words, the results presented here are largely independent of the dimension of the radiating gas (i.e., $L$). For example, given an observed light curve for a radiation transient one could modify a model to the data by, say, compensating a decrease in $L$ by a corresponding increase in $n_0$; superradiance models primarily provide a measure of $n_0 L$, but not of $n_0$ or $L$ independently \citep{Rajabi2016A,Rajabi2016B,Rajabi2017,Rajabi2019}. But we also note that, although the length scale used in our example ($L=2\times 10^{15}\,\mathrm{cm}$) might at first seem large, there is evidence for extended regions of inverted gas as large as a few hundred au in some maser-hosting regions (e.g., S255 NIRS 3; see \citealt{Moscadelli2017}).

\subsection{Transition from maser to superradiance}\label{subsec:transition}

In Sections \ref{subsec:quasisteady} and \ref{subsec:transient}, we discussed opposite limits of the Maxwell-Bloch equations where two distinct radiation processes are realized depending on the conditions relating the different time-scales characterising the system. While conditions corresponding to superradiance are very strict, those for masers seem to be more relaxed. As we show in Appendix \ref{sec:dephasing}, even when the temporal variations of the inversion density occur on time-scales shorter than $T_1$, which is contrary to the condition specified in the quasi-steady limit discussed in Section \ref{subsec:quasisteady}, non-coherent amplification or a maser can still ensue as long as dephasing processes are dominant. In real systems, the dephasing time-scale $T_2$ is usually consistently shorter than the non-coherent relaxation time-scale $T_1$ \citep{Benedict1996} and the sufficient criteria for the realisation of a maser in an inverted system is the slow variation in polarisation amplitude in relation to $T_2$. That is, whenever 
\begin{equation}
\frac{\partial P^+}{\partial\tau} \ll \frac{P^+}{T_2}  \label{eq:MaserCondition} 
\end{equation}
the evolution of coherence is greatly hampered in the system. This implies that masers can be realized beyond the strict quasi-steady state limit discussed in Section \ref{subsec:quasisteady}, the condition given by equation (\ref{eq:MaserCondition}) being sufficient. Superradiance, on the contrary, is exclusive to the fast transient limit in the sense that it requires rapid temporal variations for both the polarisation amplitude and the inversion density relative to $T_1$ and $T_2$. The question then arises as to how these two distinct processes (i.e., masing action/stimulated emission and superradiance) are linked in the evolution of a radiating system. 

To provide an answer to this question, we generalise the analysis presented in Section \ref{subsec:transient} by re-introducing the non-coherent time-scales $T_1$ and $T_2$ in the Maxwell-Bloch equations but restricting them to be equal. That is, we set a unique non-coherent time-scale $T^\prime=T_1=T_2$. This simplifying condition allows us to once again formulate a solution based on the sine-Gordon equation (see equation (\ref{eq:theta2ndOrder})), but with an accordingly adapted parameter
\begin{equation}
    q = 2\sqrt{\frac{z\tau^\prime}{LT_\mathrm{R}}}, \label{eq:q_tau'}
\end{equation}
with $\tau^\prime=T^\prime\left(1-e^{-\tau/T^\prime}\right)$ \citep{Rajabi2016A}. An example for a corresponding intensity curve is shown in Figure \ref{fig:I/Ip}, where we set $T^\prime=800\,T_\mathrm{R}$ (light/cyan curve). As can be seen from a comparison with the ideal superradiance case also presented in the figure, the presence of non-coherent relaxation and dephasing processes brings a damping to the system's response, resulting in a decrease in the peak intensity as well as in the number of bursts. This behaviour is increasingly accentuated as $T^\prime$ becomes smaller, until coherence is inhibited when $T^\prime \lesssim\tau_\mathrm{D}$. Combining this condition with equations (\ref{eq:TR}) and (\ref{eq:tau_D}) allows us to define a critical value for the column density of the population inversion 
\begin{equation}
\left(n_0L\right)_{\mathrm{crit}} \approx \frac{2\pi}{3\lambda^2}\frac{\tau_\mathrm{sp}}{T^\prime}\left|\ln\left(\frac{\theta_0}{2\pi}\right)\right|^2\label{eq:superradianceCondition2}     
\end{equation}
that must be met or exceeded for superradiance to ensue. That is, a transition from a radiation mode characterised by the masing action to one supporting superradiance will happen through a transient regime whenever the column density of the population inversion is somehow increased such that $\left(n_0L\right)\gtrsim \left(n_0L\right)_{\mathrm{crit}}$ \citep{Rajabi2016B,Rajabi2017}. Although this condition, as well as equation (\ref{eq:superradianceCondition2}), strictly applies only when $T_1=T_2$, the notion of a critical level for the column density of the population inversion is applicable to more general conditions \citep{Gross1976}. For example, such a threshold will also be observed for cases when $T_1\neq T_2$, as will be considered in the next section (and a corresponding relation is derived in Appendix \ref{sec:lethargic}).

Finally, we note that the ringing effect is more noticeable when the population inversion is established on a short time-scale compared to $T_\mathrm{R}$. This is clearly the case for the two examples with instantaneous inversion shown in Figure \ref{fig:I/Ip}. Although very fast inversion is possible in laboratory settings \citep{Skribanowitz1973}, such conditions are less likely in the interstellar medium and we should therefore not expect the ringing effect to be as clearly defined in astronomical environments. This has already been borne out by observations \citep{Szymczak2018b,Rajabi2019} and will also be apparent in the numerical examples to be discussed below.  

\section{numerical examples and discussion}\label{sec:examples}

To better understand what dictates the radiation mode of an inverted system in a given situation and how the transition between the maser and superradiance regimes takes place, we examine two separate cases in what follows. Because we will now have $T_1\neq T_2$ we must abandon the simple sine-Gordon solution for the Maxwell-Bloch equations, which are now simultaneously solved numerically with a fourth-order Runge-Kutta method for a system of methanol molecules enclosed in a cylinder of the same dimensions as in Sections \ref{subsec:transient} and \ref{subsec:transition} (see \citealt{Mathews2017}, \citealt{Houde2019} and \citealt{Rajabi2019} for details). That is, we again set the cylinder's length and radius to $L = 2 \times 10^{15}\,\mathrm{cm}$ and $w = 5.4 \times 10^7\,\mathrm{cm}$, respectively. The system is inverted either a) instantaneously at $\tau = 0$ (Section \ref{subsec:step}) or b) through a pumping pulse (Section \ref{subsec:pulse}). The pumping source for the population inversion is of the form 
\begin{equation}
\Lambda_{\mathrm{n}}\left(z,\tau\right) = \Lambda_{\mathrm{0}} + \frac{\Lambda_{\mathrm{1}}}{\cosh^{2}\left[\left(\tau-\tau_0\right)/T_\mathrm{p}\right]},\label{eq:pump}
\end{equation}
and is propagating along the symmetry axis of the cylinder (see equation (\ref{eq:dN/dt}) and \citealt{Rajabi2019}). In both cases considered a constant pump rate $\Lambda_{\mathrm{0}}$ is present throughout the evolution of the system, while $\Lambda_{\mathrm{1}}$ is the amplitude of the pump pulse centered at the retarded time $\tau_0$ and also characterised by a time-scale $T_{\mathrm{p}}$ (for Section \ref{subsec:pulse}). The particular profile of the pump pulse is not central to our discussion. The non-coherent relaxation and dephasing time-scales are set to $T_1=1.64\times10^7$~s and $T_2=1.55\times10^6$~s throughout (i.e., $T_1\simeq11\,T_2$). Once again, the methanol molecules interact with one another through their common radiation field at $6.7\,\mathrm{GHz}$. 

In order to realize our stated goals we will compare the radiation intensity, the population inversion density and the polarisation amplitude between the steady-state (maser) and transient (superradiance) responses of our system of radiating molecules. This will allow us to investigate the behaviours of these different parameters as a function of the radiation processes at play.  

\subsection{Response to a fixed initial population inversion level (instantaneous inversion)}\label{subsec:step}

For our first example we return to the case of an instantaneous inversion treated in Sections \ref{subsec:transient} and \ref{subsec:transition}, but with two important differences. As mentioned before the non-coherent relaxation and dephasing time-scales are now unequal (i.e., $T_1=1.64\times10^7$~s and $T_2=1.55\times10^6$~s) while, although the initial population inversion level is still set to $n_0=3.3\times 10^{-12}\,\mathrm{cm^{-3}}$, a constant inversion pump level $\Lambda_0=n_0/\left(2T_1\right)$ is present for $\tau>0$ at all positions $z$ along the system. This constant pump term keeps the population inversion level to $n_0$ when no or little radiation emerges at the end-fire ($z = L$) of the system (e.g., for the unsaturated maser regime). No inversion pump pulse is applied, i.e., $\Lambda_1=0$ (see equation (\ref{eq:pump})). Although the case of an instantaneous inversion is somewhat idealised, it will allow us to more clearly see the behaviour of the system in the different radiation modes. 
The initial population inversion level $n_0$ was chosen to ensure that the critical threshold for the initiation of superradiance will be met approximately halfway along the cylinder, i.e., at $z\simeq L/2$. It will thus be instructive to compare the radiation intensity and the different parameters characterising the state of the system at a position upstream of $z=L/2$ and at the end-fire (i.e., at $z=L$). 
\begin{figure*}
    \centering
    \begin{minipage}[t]{.48\textwidth}
    \includegraphics[width=\columnwidth]{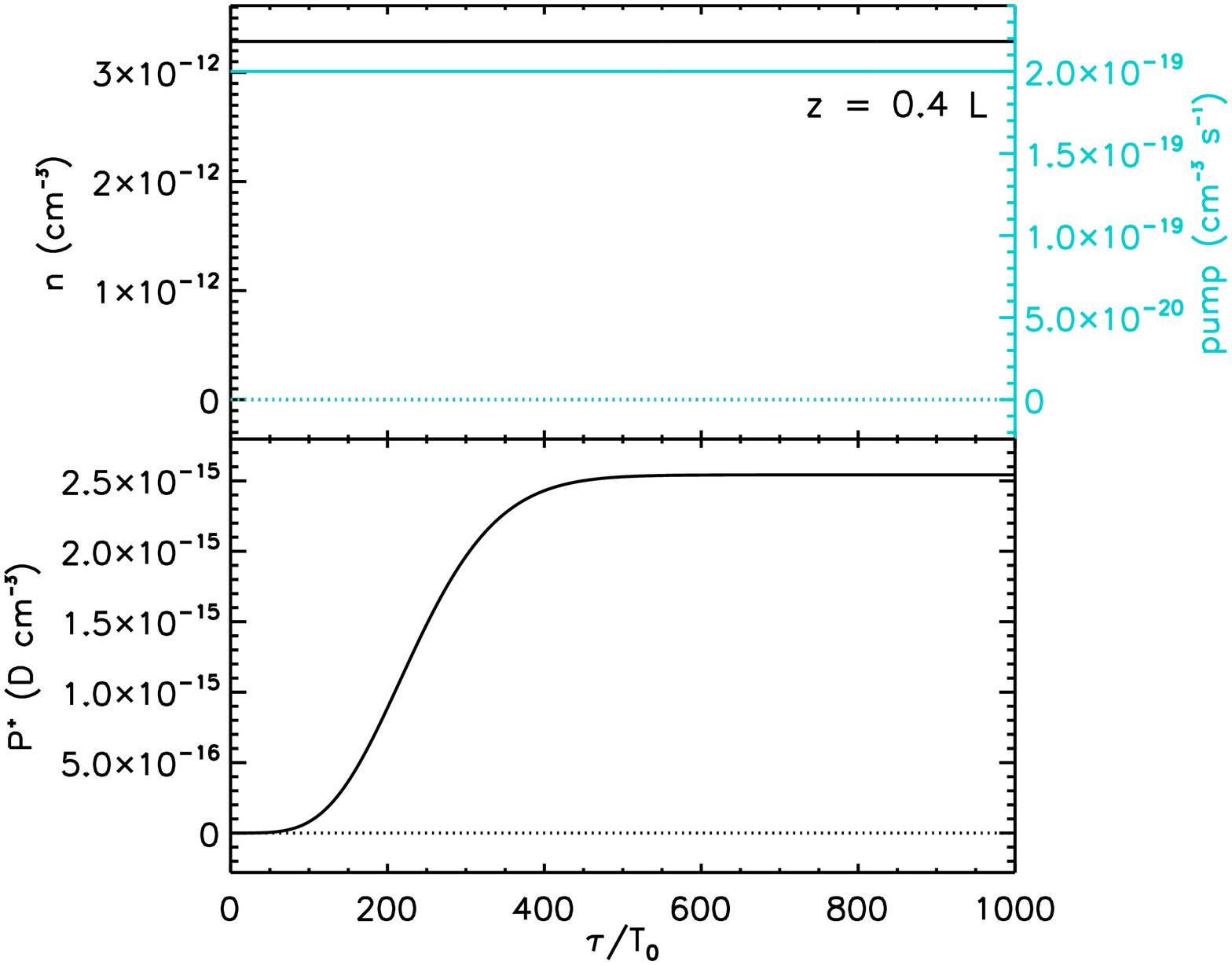}
    \end{minipage}\qquad
    \begin{minipage}[t]{.48\textwidth}
    \includegraphics[width=\columnwidth]{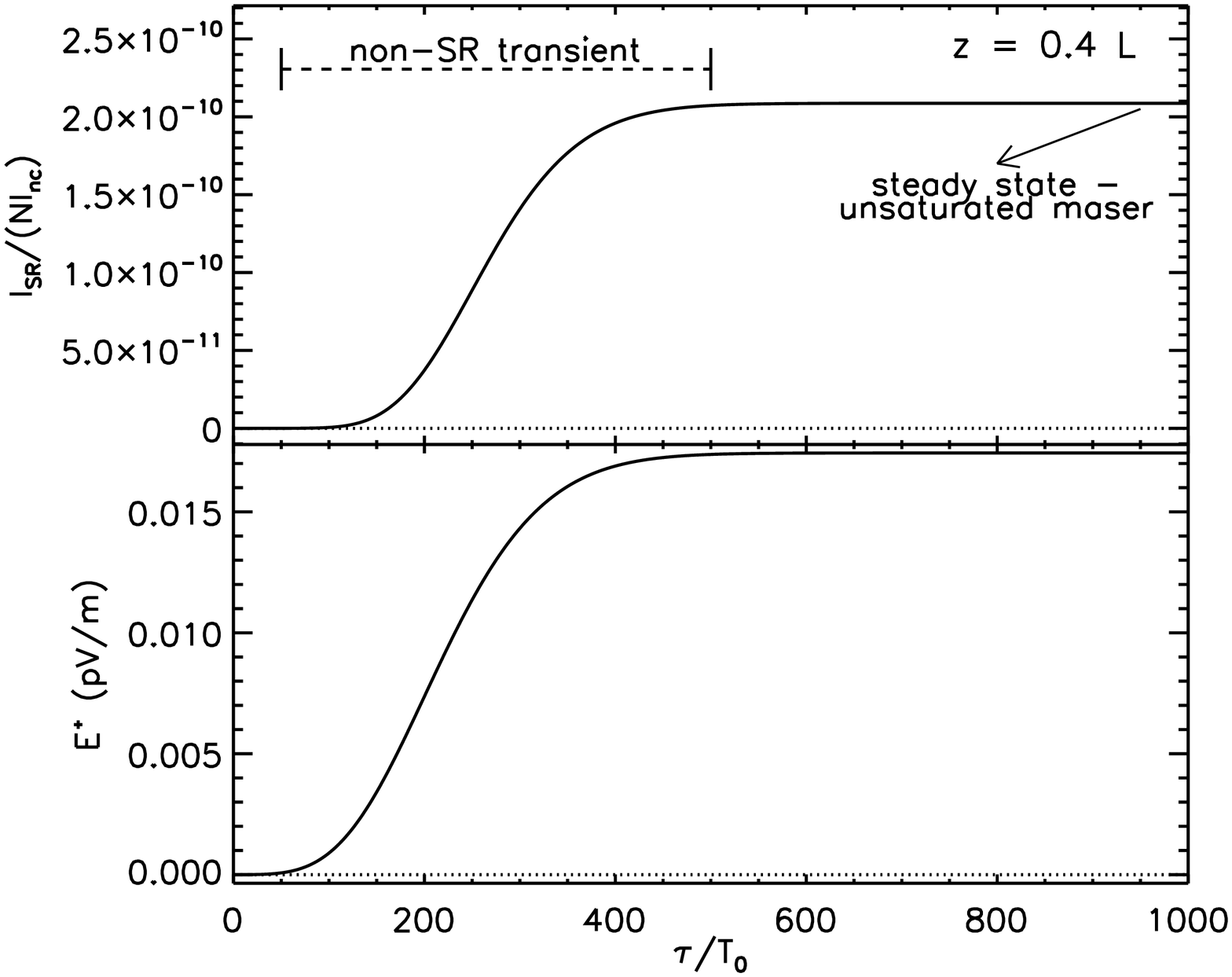}
    \end{minipage}
    \caption{Temporal evolution at $z=0.4L$ for the population inversion density $n$ and polarisation amplitude $P^+$ (left panel), as well as the normalised intensity $I_\mathrm{SR}/\left(NI_\mathrm{nc}\right)$ and electric field amplitude $E^+$ (right panel) as a function of the retarded time in units of $T_0=1\times 10^5$~s. The dimensions of the system are the same as in Figure \ref{fig:I/Ip} but here $T_1=164\,T_0$ and $T_2=15.5\,T_0$. The system is instantaneously inverted at a level of $n_0=3.3\times 10^{-12}\,\mathrm{cm}^{-3}$ (top left; dark/black curve, left vertical scale) at $\tau=0$ for all $z$ by a constant pump (top left; light/cyan curve, right vertical scale). At this position in the system the inverted column density is below the critical threshold. This accounts for the low polarisation amplitude (i.e., $P^+\ll n_0 d$), which in turn implies a very weak level of coherence, a smooth (non-superradiance) transient in intensity (top right) and steady-state unsaturated maser regime. The electric field amplitude behaviour mimics that of the intensity.} 
    \label{fig:unsaturated_step}
\end{figure*}

We present in Figure \ref{fig:unsaturated_step} a set of graphs for the population inversion density and the polarisation amplitude on the left, and the intensity and the electric field on the right. These were all obtained at $z=0.4\,L$ for $\tau\geq 0$, a position where the column density of the inverted population is below the critical level needed for superradiance. The retarded time is shown in units of $T_0=1\times10^5$~s (note that $T_\mathrm{R}=4\times10^4$~s as in Sections \ref{subsec:transient} and \ref{subsec:transition}). Focusing first on the left side of the figure we observe that the population inversion density (top panel; dark/black curve using the vertical scale on the left) remains constant throughout at $n=n_0$ because of the action of the constant pump level (light/cyan curve, using the vertical scale on the right). The bottom (left) panel shows that the evolution of the polarisation amplitude $P^+$ goes through a smooth transient phase (i.e., no overshoot is observed) between $100\lesssim\tau/T_0\lesssim 400$, after which period it settles to a steady-state value $P^+\sim 10^{-15}\,\mathrm{D~cm}^{-3}$. Given the electric dipole moment associated with the 6.7~GHz transition ($d\simeq 0.7$~D) and the population inversion density this value for the polarisation is very weak since $P^+\ll n_0 d$. This is a signature of a system exhibiting a very low level of coherence. That is, the individual molecular dipoles do not oscillate with a well-defined and organised phase relationship. 

Moving to the right panel of Figure \ref{fig:unsaturated_step} we show the corresponding radiation intensity (top) normalized to $N$-times that expected from $N$ individual molecules emitting spontaneously and randomly in a non-coherent manner. That is, if $I_1$ is the intensity from one molecule, the non-coherent emission from $N$ such molecules is $I_\mathrm{nc}=N I_1$, while we normalized the radiation intensity from our system to $N I_\mathrm{nc}$ \citep{Rajabi2016A,Rajabi2016B}. This is because in the ideal limit a so-called small sample superradiance system (for which $L\ll\lambda$) can be shown to radiate with an intensity $I\sim N^2 I_1$ \citep{Dicke1954}. As we can see, the radiation intensity also exhibits a smooth transient regime between $100\lesssim\tau/T_0\lesssim 400$, which, for reasons to become clear later on, we label ``non-SR transient'' (``SR'' standing for ``superradiance''). Accordingly with the low polarisation level, the steady-state radiation intensity level $I_\mathrm{steady}\sim 10^{-10} N I_\mathrm{nc}$ does not exhibit a sustained level of coherence (once again $N=6\times 10^{19}$ for this system). As we will soon see, this radiation regime corresponds to that of an unsaturated maser. The corresponding electric field shown in the bottom panel exhibits a similar response as the intensity, as could be expected.  
\begin{figure*}
    \centering
    \begin{minipage}[t]{.48\textwidth}
    \includegraphics[width=\columnwidth]{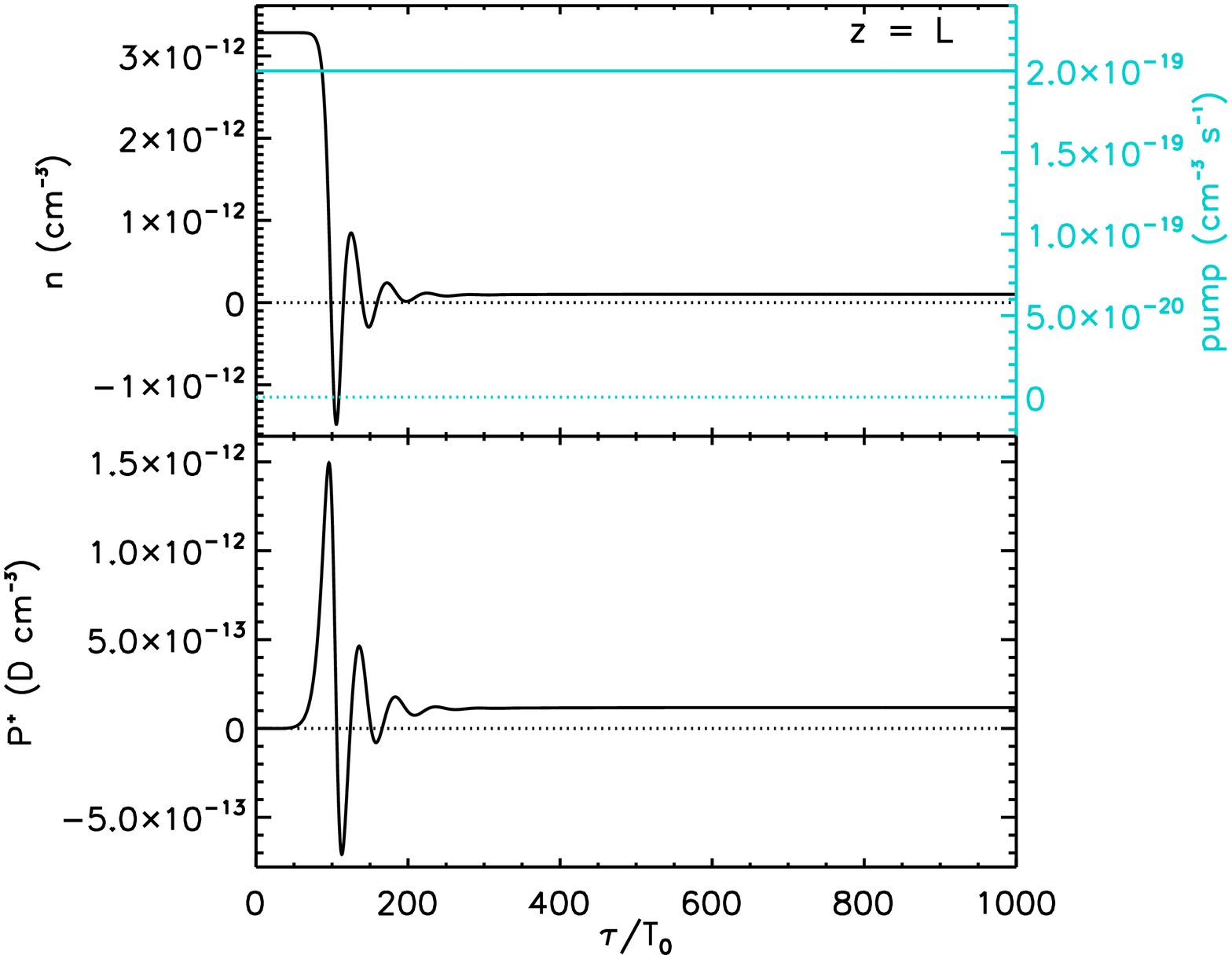}
    \end{minipage}\qquad
    \begin{minipage}[t]{.48\textwidth}
    \includegraphics[width=\columnwidth]{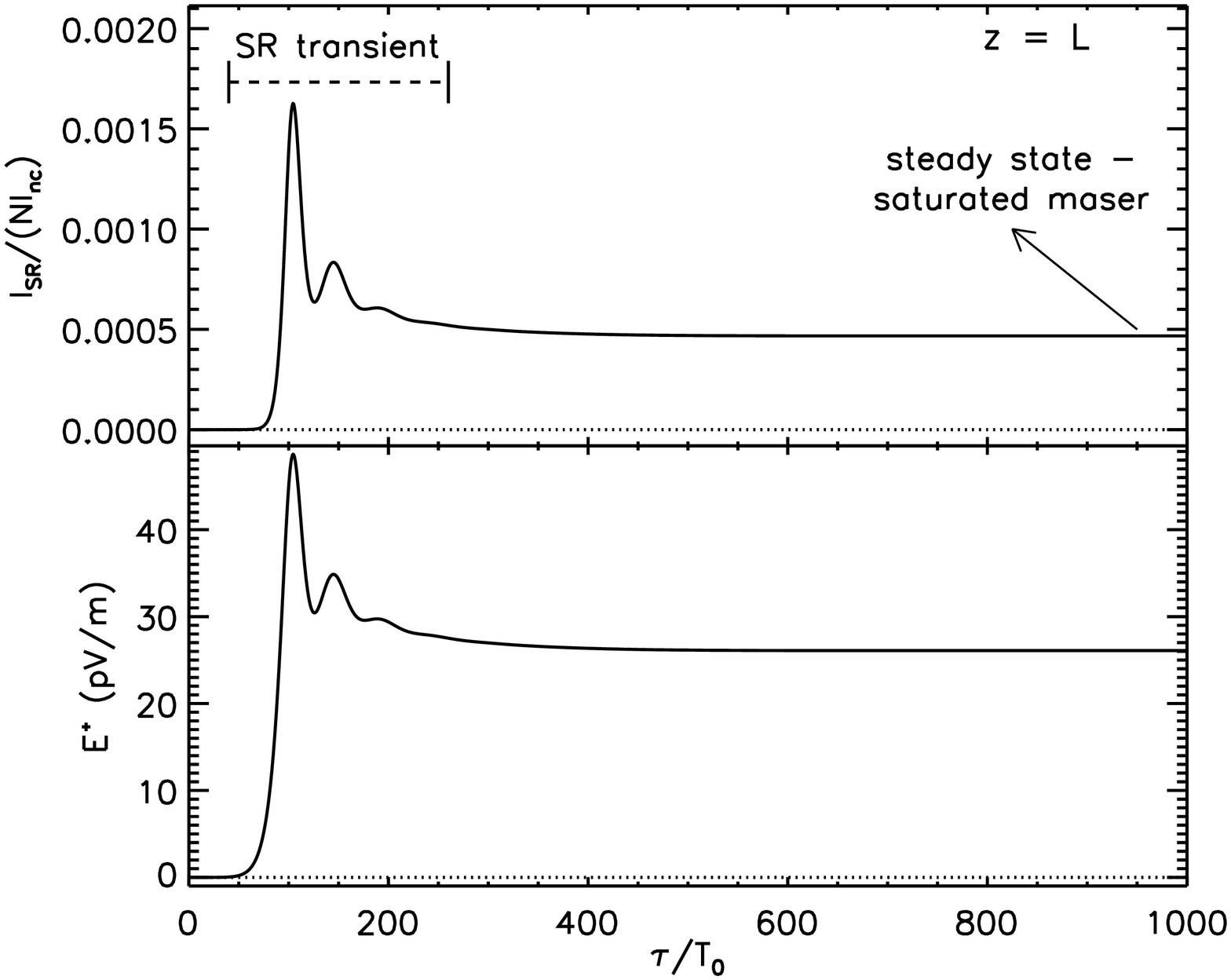}
    \end{minipage}
    \caption{Temporal evolution at the end-fire ($z = L$) for the same system and parameters as in Figure \ref{fig:unsaturated_step}. At this position in the sample the inverted column density exceeds the critical threshold. Accordingly, the inverted density and the polarisation amplitude exhibit strong oscillations during the transient regime. Notably, the polarisation peak is very high (i.e., $P^+\approx n_0 d$) resulting in a strong superradiance (SR) intensity. The steady-state regime is that of a saturated maser.}
    \label{fig:saturated_step}
\end{figure*}

The situation is drastically different at the end-fire ($z = L$) of the system, where the column density of the inverted population exceeds the critical level. This can be assessed from Figure \ref{fig:saturated_step} where we display the temporal behaviour of the same parameters at $z=L$. In the left panel of the figure, we observe strong oscillations in both $n$ (top) and $P^+$ (bottom) starting at $\tau/T_0\simeq 100$. In the process the population inversion density abruptly drops from its initial value $n_0$ to reach negative values, where more molecules are in the lower state than in the upper state, before eventually settling to a steady-state nearing a zero inversion level. We note that the peak value of the polarisation reaches $\sim n_0 d$ (within a factor of two), which is an indication of a high level of coherence in the system as the transient regime is established. This implies that almost all molecular dipoles are now oscillating with a well-defined phase relationship. The polarisation level drops by about an order of magnitude once the steady-state regime is attained. 

In the right panel of Figure \ref{fig:saturated_step}, we can see a clear overshoot in intensity (top) at the start of the transient regime, where a peak intensity $I_{\mathrm{p}} \simeq 1.5 \times 10^{-3} N I_\mathrm{nc}$ is reached after a time delay $\tau_{\mathrm{D}}\sim 100\,T_0$. This overshoot, indicating a fast release of energy stored in the system, is a signature of superradiance, which is established when the polarisation and coherence levels are high enough. During the transient superradiance regime, a clear ringing effect can be seen in the intensity profile where the first burst of radiation is followed by two smaller ones. Because of the presence of the constant pump level the intensity then gradually settles into a steady-state regime of intensity approximately 30\% of the peak level at $\tau \gtrsim 500\,T_0$. The corresponding electric field profile for the evolution of the system, shown in the bottom of the right panel, also exhibits a similar oscillatory pattern plateauing out into a steady state. The ringing of the intensity (and the electric field) is the result of absorption and re-emission at the end-fire ($z = L$) of radiation emanating from positions at $z<L$, a phenomenon also reflected in the temporal variations of inverted population density at $z = L$ (i.e., returning to positive $n$ values after dropping into negative ones; \citealt{Benedict1996}). The steady state phase of the evolution shown in Figure \ref{fig:saturated_step} corresponds to the saturated maser regime, as will soon be discussed.   

A comparison of the system's response when its inversion level is below ($z = 0.4L$; Figure \ref{fig:unsaturated_step}) and above ($z = L$; Figure \ref{fig:saturated_step}) the critical threshold reveals different evolutionary phases depending on the degree of coherence achieved within the molecular population. When the column density of the inverted population is below the critical level, we observe nearly non-coherent (i.e., non-SR) transient and (unsaturated maser) steady state regimes, whereas the high level of coherence exhibited when above criticality allows the realisation of superradiance and the subsequent settling into a steady state saturated maser. 

\begin{figure}
    \centering
    \includegraphics[width=\columnwidth]{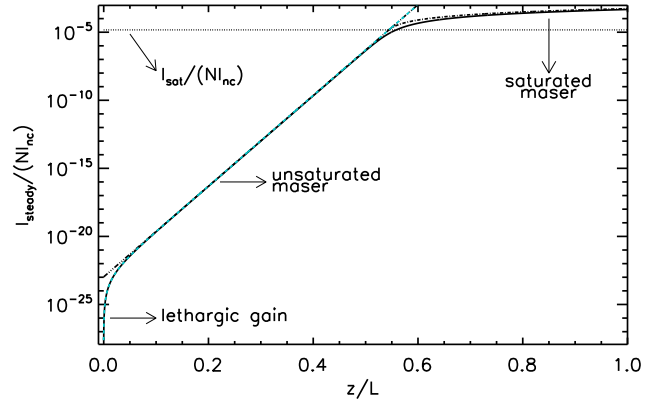}
    \caption{Steady-state (maser) intensity at the end-fire ($z = L$) normalised to $NI_{\mathrm{nc}}$. The two well-known maser regimes are indicated: the unsaturated (exponential growth) and saturated (linear growth) domains. These two modes meet at the critical length $z_\mathrm{crit}/L\approx 0.54$, where the intensity neighbours the saturation intensity $I_\mathrm{sat}$ (shown with the horizontal dotted line). As well, at $z<0.01$ we observe the fast rise in intensity in the so-called lethargic gain regime. Curves for the theoretical expressions pertaining to the three regimes are also shown: equations (\ref{eq:IntensityMaserWeak}) (with equation (\ref{eq:I0_main-text})) and (\ref{eq:IntensityMaserStrong2}) for the unsaturated (triple-dot broken line) and saturated masers (dot-broken curve), respectively, while that for the lethargic gain (equation (\ref{eq:E+_lethargic})) was plotted in light/cyan (broken curve), as it is not distinguishable it from that of the numerical solution (dark/black solid curve) and merges with the theoretical line for the unsaturated maser regime at $z/L\gtrsim0.1$.} 
    \label{fig:Imaser}
\end{figure}

To get a more complete picture of the behaviour of the system, we will investigate the steady state and transient regimes of the intensity, population inversion density and polarisation as a function of distance $z$ along the sample. But before we do so, we first focus on the steady state intensity to clearly establish its correspondence to maser radiation. In Figure \ref{fig:Imaser}, we show the normalised steady state intensity $I_{\mathrm{steady}}/\left(NI_{\mathrm{nc}}\right)$, measured at $\tau=1000\,T_0$, as a function of the fractional distance $z/L$, using a logarithmic scale for the vertical axis. Although, as expected, the intensity grows with distance as the inverted column density increases, the rate of change depends on $z/L$ and its position in relation to $z_{\mathrm{crit}}/L$ at which the critical inverted column density is reached. As mentioned earlier, in this example we set $z_{\mathrm{crit}} \simeq 0.5 L$. A drastic change in behaviour is observed around this critical value, where a rapid increase in the steady state intensity relative to $NI_{\mathrm{nc}}$ below $z/L\approx0.5$ gradually settles into a flatter response above it. This coincides with what one expects for the transition from an unsaturated to a saturated maser when the intensity exceeds the saturation intensity $I_{\mathrm{sat}}$ (shown with the horizontal dotted line in the figure; see equation (\ref{eq:Is}) and Figure 1.6 in \citealt{Gray2012}). As discussed in Section \ref{subsec:quasisteady}, the intensity of an unsaturated maser grows exponentially with $z$, similar to the trend seen in Figure \ref{fig:Imaser} for $0.1\lesssim z/L\lesssim 0.5$, while the saturated maser intensity increases linearly with $N$ and accordingly with position for a homogeneous gas (for $z\gtrsim 0.6$ in Figure \ref{fig:Imaser}). These limits are accordingly identified in the figure. 

Interestingly, we note that near the front-end of the sample for $z/L\lesssim 0.01$ the intensity is in the so-called lethargic gain regime where it scales as $I_0\left(e^{\alpha z/2}-1\right)^2\propto z^2$ (see equation (\ref{eq:E+solution3}) in Appendix \ref{sec:lethargic}). The intensity growth in this range is thus coherent and is significantly different from that in the unsaturated maser regime at larger $z/L$ (see equation (\ref{eq:IntensityMaserWeak})). The behaviour of the system in this regime cannot be understood from the quasi-steady state limit used in Section \ref{subsec:quasisteady} or Appendix \ref{sec:dephasing}, since for small $z/L$ the relative change in polarisation with time due to internal quantum fluctuations (i.e., spontaneous emission) can be significant (e.g., see equations (\ref{eq:Ncosine})-(\ref{eq:Nsine}) for $\theta\ll 1$). One must then resort to an analysis grounded in the coherent regime of the Maxwell-Bloch equations. The details are given in Appendix \ref{sec:lethargic}. We note, however, that this phase is unlikely to be realized in astronomical media. This is because the seed provided by background radiation will usually greatly exceed that due to the aforementioned internal quantum fluctuations responsible for initiating the lethargic gain regime. For example, we have $I_0\approx10^{-18}I_\mathrm{sat}$ in the present case (from Figure \ref{fig:Imaser}) while a typical value for background radiation of astrophysical masers is on the order of $10^{-7}I_\mathrm{sat}$ \citep{Gray2012}.

In Figure \ref{fig:Imaser} we have also plotted curves for the theoretical expressions pertaining to the three regimes. We used equation (\ref{eq:IntensityMaserWeak}) for the unsaturated maser (broken line) with (see Appendix \ref{sec:lethargic}) 
\begin{equation}
    I_0 = \frac{c\epsilon_0}{2}\left(\frac{\hbar\theta_0}{2dT_2}\right)^2,\label{eq:I0_main-text}
\end{equation}
and
\begin{equation}
    I(z) = \frac{\hbar\omega_0 n_0}{8 T_1}\left(z-z_\mathrm{crit}\right)+I_\mathrm{sat},\label{eq:IntensityMaserStrong2}
\end{equation}
for the saturated maser regime (dot-broken curve) with $z_\mathrm{crit}\simeq 0.54L$ from equation (\ref{eq:nLcrit}). A curve was plotted in light/cyan (broken curve) for the lethargic gain, using equation (\ref{eq:E+_lethargic}), as it is practically impossible to distinguish it from that of the numerical solution (solid curve) and merges with the theoretical line for the unsaturated maser regime at $z/L\gtrsim0.1$.  

\begin{figure}
    \centering
    \includegraphics[width=\columnwidth]{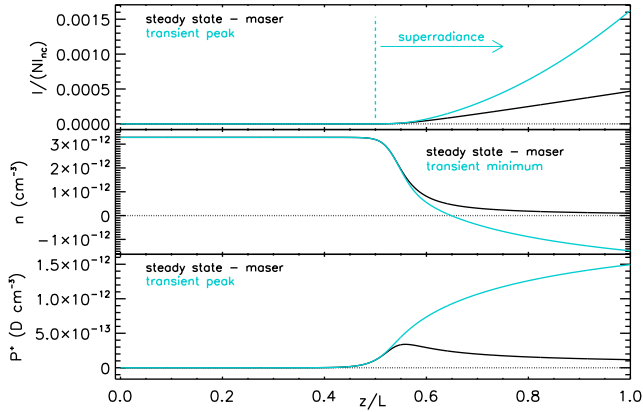}
    \caption{Comparison between the transient (light/cyan curves) and steady-state (dark/black curves) normalised intensities (top), population inversion densities (middle) and polarisation amplitudes (bottom) as a function of the position $z$ in the system in units of its length $L$. The superradiance regime is established at $z/L\approx 0.5$ where the critical inverted column density is reached, the system then transitions between the unsaturated and saturated maser regimes and the steady state intensity neighbours the maser saturation intensity $I_\mathrm{sat}$. Beyond that point the superradiance peak intensity scales as $z^2$ while the maser's is linear with position ($\propto z$). The superradiance inversion minimum level reaches more negative values with increasing $z$ (because of the strong oscillations during the transient regime) while the maser inversion asymptotically tends to zero. Finally, we also note that the superradiance polarisation amplitude (and the level of coherence) increases with distance, contrarily to the maser's which diminishes as the steady-state intensity increases.}
    \label{fig:stats-step}
\end{figure}

In Figure \ref{fig:stats-step}, the normalised intensity $I/\left(NI_{\mathrm{nc}}\right)$ (top panel), polarisation amplitude (middle panel) and inverted population density (bottom panel) at different fractional distances $z/L$ are compared between the steady-state (dark/black curves) and transient (light/cyan curves) regimes. As already discussed, the steady state intensities below and above the critical point (i.e., $z_{\mathrm{crit}}/L \approx 0.5$ for this system) correspond to unsaturated and saturated maser intensities, respectively. The transient regime for $z$ below the critical point is expected to manifest characteristics of a nearly non-coherent evolution while for $z>z_\mathrm{crit}$, when the threshold inverted column density is reached, it is expected to trace the behaviour of superradiance. The comparison of the peak value of the intensity in the transient regime with the steady-state intensity at each $z$ verifies these statements. More precisely, in the top panel of the figure for $z/L>z_{\mathrm{crit}}/L$ the normalised peak transient intensity (light/cyan curve) exhibits a quadratic growth with distance (i.e., it is proportional to $N^2$ for an homogeneous medium) that exceeds the corresponding linear behaviour (i.e., proportional to $N$) of the saturated maser intensity (dark/black curve). As a result, the separation between the two curves increases as we move to further positions along the sample. As discussed in Section \ref{subsec:transient} the scaling of the peak intensity with $N^2$ is a characteristic of superradiance (see equation (\ref{eq:PeakIntensitySR})); we accordingly labeled the corresponding curve in the figure. For $z/L<z_{\mathrm{crit}}/L$, the peak transient intensity, on the contrary, mimics that of the steady-state intensity in the unsaturated maser regime, as expected for a system exhibiting a very weak level of coherence. 

The inverted population density depletion is also investigated below and above $z_\mathrm{crit}$ in the steady state and transient regimes (middle panel of Figure \ref{fig:stats-step}). Similarly to what is seen in Figure \ref{fig:unsaturated_step} for $z = 0.4L$, the inverted population density in the steady-state regime (dark/black curve) remains constant for $z/L<z_{\mathrm{crit}}/L$. This is because the loss of inversion through the unsaturated maser action and the non-coherent relaxation processes is compensated by the constant inversion pump rate $\Lambda_0$ (see equation (\ref{eq:pump})). For $z/L>z_{\mathrm{crit}}/L$, the depletion of the inverted population density becomes significant in the saturated maser regime as $n$ starts dropping abruptly for $z \gtrsim 0.5L$ and asymptotically approaches zero as $z\rightarrow L$ at the end-fire ($z = L$). It should be noted that in the steady state regime $n>0$ at all $z$. In the transient regime, where for a meaningful comparison we plotted the minimum value of $n$ at every position, at $z/L>z_{\mathrm{crit}}/L$ the inverted population density is depleted at a faster rate than that of the saturated maser. This is consistent with the observations previously made for the intensities shown in the top panel of the figure (i.e., fast depletion in the inverted population density implies a higher photon emission rate and radiation intensity). A salient feature of the population inversion density curve for the transient regime is the realisation of negative values for $n$ as we approach the end-fire ($z = L$). More generally, rapid fluctuations between positive and negative values, as seen in the top left panel of Figure \ref{fig:saturated_step}, is a characteristic of superradiance. 

Another interesting parameter to examine is the polarisation amplitude, which gives a measure of coherence during the evolution of the system. Accordingly, we show in the bottom panel of Figure \ref{fig:stats-step} the polarisation amplitude for the steady state and transient-peak values as a function of $z/L$. As can be seen in the figure, the polarisation levels corresponding to the peak transient (light/cyan curve) and steady state (dark/black curve) intensities remain close to zero for $z/L\lesssim0.5$. This indicates the nearly non-coherent behaviour of the system when $z<z_{\mathrm{crit}}$. Beyond $z_{\mathrm{crit}}/L$, a macroscopic polarisation quickly emerges in the transient regime, peaking at approximately $1.5\times 10^{-12}\,\mathrm{D~cm}^{-3}$ at the end-fire (i.e., $z = L$). This corresponds to close to a three orders of magnitude increase from the polarisation level attained at $z=0.4L$ upstream of the critical point shown in Figure \ref{fig:unsaturated_step}. As stated before, this level is close to $n_0d$, the maximum polarisation that could be reached, and coincides with the fast build-up of coherence and the subsequent fast release of energy observed in the SR-transient part of the corresponding intensity curve (i.e., right panel in Figure \ref{fig:saturated_step}). It should be also noted that the polarisation level grows slightly above $z_\mathrm{crit}$ in the saturated maser response, which indicates the presence of some degree of coherence in that regime. But as we move along the sample to larger $z/L$ values, while the superradiance polarisation amplitude keeps increasing, that of the saturated maser asymptotically tends to zero. In other words, the macroscopic polarisation keeps building up in the system as coherence develops through the superradiance regime, while it tends to fade away as the saturated maser gains intensity. 

\begin{figure}
    \centering
    \includegraphics[width=\columnwidth]{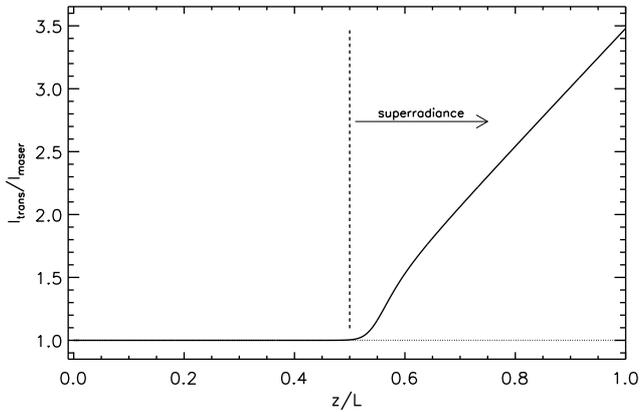}
    \caption{Ratio of the transient to maser intensities $I_{\mathrm{trans}}/I_{\mathrm{maser}}$ as a function of the normalised distance $z/L$. We clearly see that the superradiance regime is established at the critical length $z_\mathrm{crit}/L\approx 0.5$, where the ratio starts increasing linearly with $z$. The critical length also coincides with the location in the system where the steady-state regime transitions between the unsaturated and saturated maser domains.}
    \label{fig:ratio-step}
\end{figure}

To quantify how the superradiance and maser intensities scale in relation to one another we show $I_{\mathrm{trans}}/I_{\mathrm{maser}}$, the ratio of the transient (peak) to the maser (steady state) intensities, as a function fractional distance $z/L$ in Figure \ref{fig:ratio-step}. We again note the significant change in behaviour around $z_{\mathrm{crit}}/L \approx 0.5$ in the figure. As discussed earlier, for $z<z_{\mathrm{crit}}$ the system is in a nearly non-coherent regime for which there is no overshoot in the transient response and thus $I_{\mathrm{trans}}/I_{\mathrm{maser}} = 1$. Above $z_\mathrm{crit}$ the transient superradiance intensity grows quadratically with $z$ while the saturated maser does so linearly, leading to the ratio $I_{\mathrm{trans}}/I_{\mathrm{maser}}\propto z$ observed in the figure. Evidently, the gain in intensity of superradiance over the saturated maser will increase further with the length $L$ of the system. That is, if $I_{\mathrm{trans}}/I_{\mathrm{maser}} = 3.5$ at the end-fire ($z = L$) of our system ($L=2\times 10^{15}$~cm), it would be $\approx 8.5$ for a system twice as long, and so on for longer lengths.  

In the next section, we show how the gain of superradiance over a maser can be orders of magnitude larger than seen in this section when the column density of the inverted population is pushed beyond its critical level through the action of a pump pulse.  

\subsection{Response to a pulsed pump source}\label{subsec:pulse}

\begin{figure*}
    \centering
    \begin{minipage}[t]{.48\textwidth}
    \includegraphics[width=\columnwidth]{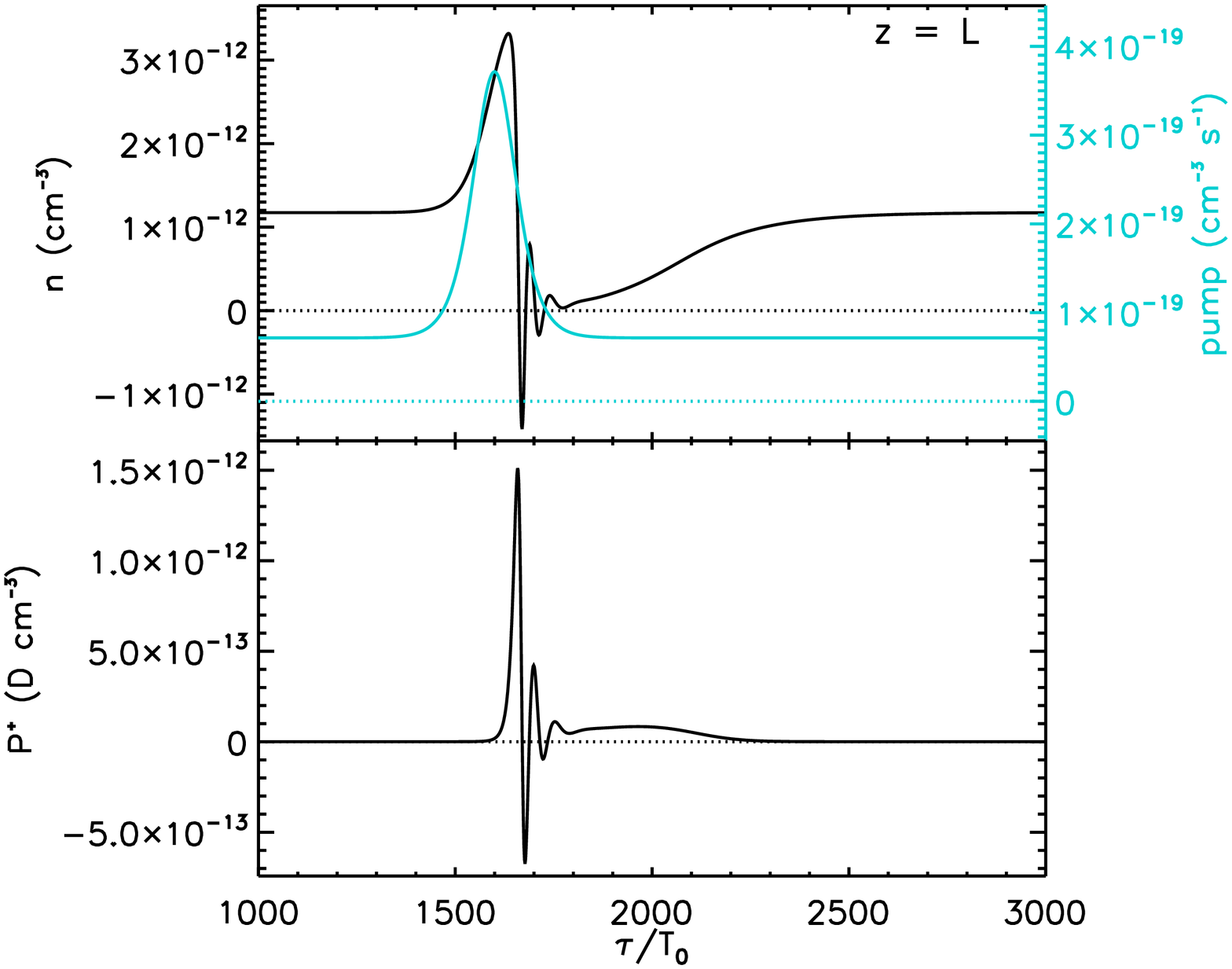}
    \end{minipage}\qquad
    \begin{minipage}[t]{.48\textwidth}
    \includegraphics[width=\columnwidth]{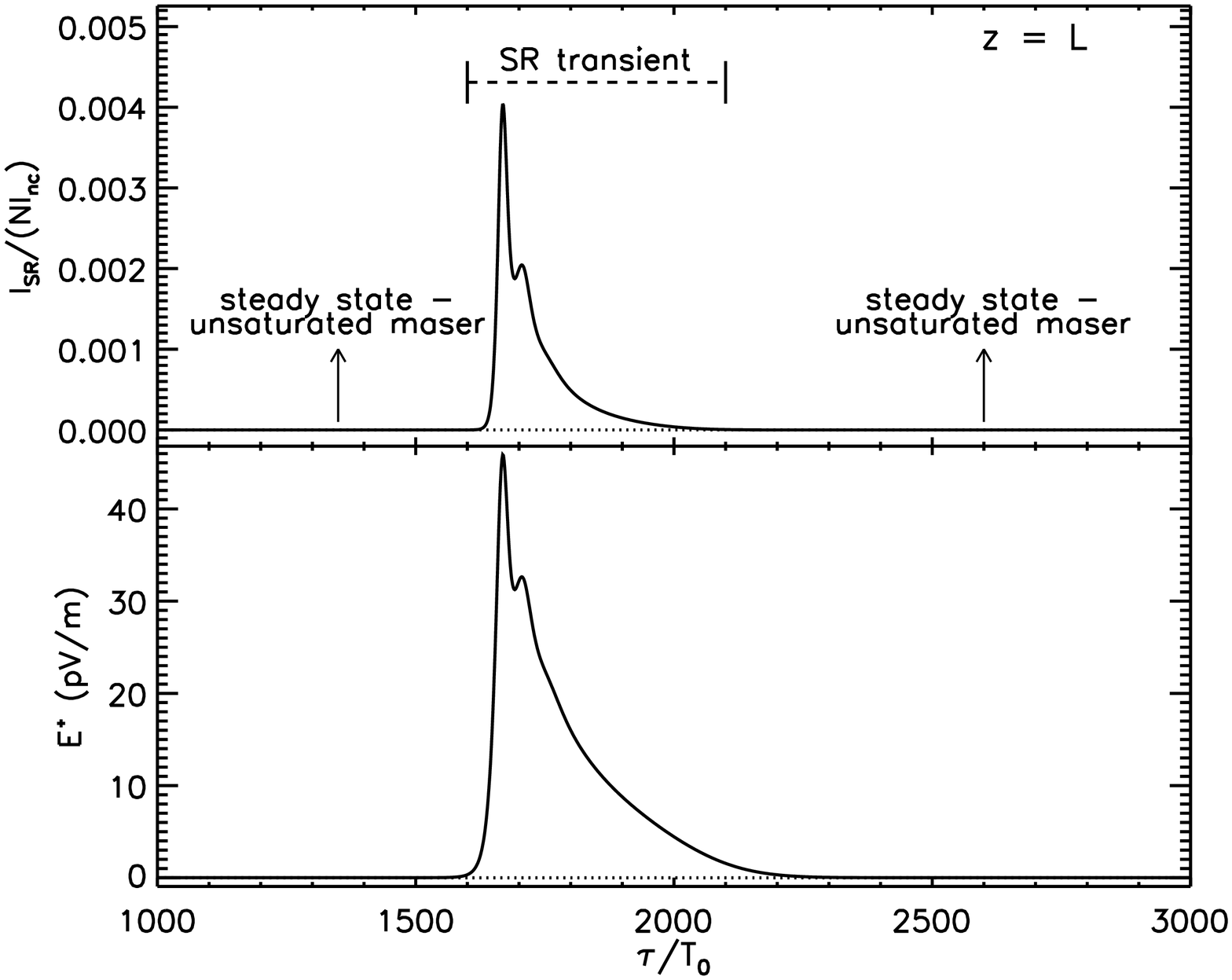}
    \end{minipage}
    \caption{Temporal evolution at the end-fire ($z=L$) for the same system and parameters as in Figures \ref{fig:unsaturated_step} and \ref{fig:saturated_step}, but for the case of a pulsed pump source. The complete pump signal (left top panel, light/cyan curve using the vertical scale on the right) also includes a constant component responsible for the initial unsaturated maser state of the system. The pump pulse brings the population inversion level (left top panel, dark/black curve using the vertical scale on the left) to the critical threshold, at which point a strong superradiance transient regime is established between $1600\leq \tau/T_0 \leq 2100$. This phase is characterised by significant oscillations in the population inversion density and polarisation amplitude (with a peak $P^+\approx n_0 d$), as well as a major increase in the radiation intensity. The electric field amplitude profile mimics that of the intensity. All parameters return to their initial values once the inversion pulse vanishes, doing so at a rate mostly set by the non-coherent relaxation time-scale $T_1$.}
    \label{fig:pulse}
\end{figure*}

The most salient feature of superradiance is observed when a pumping pulse excites an unsaturated maser (or a gas without inversion) to bring the population inversion density level past the critical superradiance threshold. Accordingly, we intend to show how in some cases increasing the population inversion level by only a small factor can move the system to a coherent regime and lead to powerful bursts of radiation through superradiance. To achieve this, we will study the output response (i.e., at $z=L$) of the same system as in Section \ref{subsec:step} to a pumping pulse of the form given by equation (\ref{eq:pump}). The parameters defining this excitation consist of the constant pump rate $\Lambda_0 = 7.15 \times 10^{-20}~\mathrm{cm}^{-3}\mathrm{s}^{-1}$, pulse amplitude $\Lambda_1 = 3.0 \times 10^{-19}~\mathrm{cm}^{-3}\mathrm{s}^{-1}$ and pulse duration $T_{\mathrm{P}} = 7.3 \times 10^{6}$~s. The pump signal is shown in the left panel of Figure \ref{fig:pulse} (top; light/cyan curve using the vertical axis on the right). 

Still in the left panel of Figure \ref{fig:pulse}, we present the temporal evolutions of the population inversion density (top; dark/black curve using the vertical axis on the left) and the polarisation amplitude (bottom). As before, we scaled the retarded time $\tau$ to $T_0=1\times 10^5$~s for the horizontal axis. As can be seen in the figure, the constant pump rate brings an initial population inversion level $n = 1.17 \times 10^{-12}~\mathrm{cm}^{-3}$, which results in an inverted column density below the critical level. This can be asserted from the near-zero polarisation level observed until $\tau/T_0=1400$ when the pulse appears. Similarly to the case discussed in Figure \ref{fig:unsaturated_step}, the state of the system during that period corresponds to an unsaturated maser. For $1400\leq \tau/T_0 \leq 1800$ the symmetric pumping pulse is applied to the system, resulting in an almost three-fold increase in $n$. This rise in the inversion level coincides with the build-up of a strong polarisation in the system, peaking at $P^+ = 1.5 \times 10^{-12}~\mathrm{D}~\mathrm{cm}^{-3}$, implying a high level of coherence in the system. In other words, the moderate increase in the population inversion was sufficient to reach the aforementioned critical level and initiate superradiance. As a result the population inversion density and the corresponding polarisation amplitude go through the fast oscillatory behaviour that is characteristic of superradiance. Both parameters return to their initial values once the inversion pulse vanishes, doing so at a rate mostly set by the non-coherent relaxation time-scale $T_1$. 

In the right panel of Figure \ref{fig:pulse}, the temporal evolution of the corresponding scaled superradiance intensity $I_{\mathrm{SR}}/\left(NI_{\mathrm{nc}}\right)$ (top) and the electric field amplitude (bottom) are shown. The plots display a fast transient regime between $1600\leq \tau/T_0 \leq 2100$, identified as ``SR transient'' on the intensity graph. The maximum intensity of $I_{\mathrm{SR}} \simeq 4 \times 10^{-3} \left(NI_{\mathrm{nc}}\right)$ coincides with that of the polarisation amplitude $P^+$ (see the left panel of the figure) and the peaking of coherence in the system, as could be expected. We also note the time delay between the appearances of the pump pulse (left panel) and the burst of radiation, which is also a characteristic feature of superradiance. For $\tau/T_0<1600$ and $\tau/T_0>2100$, the constant near-zero intensity levels correspond to the steady-state unsaturated maser regime we previously alluded to. The system, initially in that state, is brought to the superradiance regime by the pump pulse responsible for elevating the population inversion level beyond the critical threshold, and ultimately returns to the unsaturated maser regime once the stored energy is released and the initial inversion level is restored through the action of the constant pump level. The evolution of the electric field amplitude exhibits a pattern similar to that of the scaled intensity. 
\begin{figure}
    \centering
    \includegraphics[width=\columnwidth]{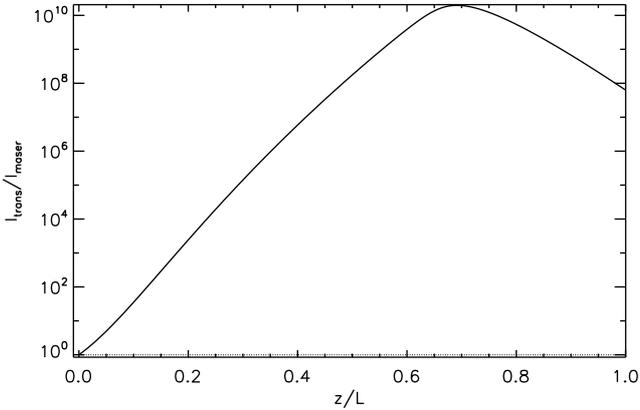}
    \caption{Ratio of the transient to maser intensities $I_{\mathrm{trans}}/I_{\mathrm{maser}}$ as a function of the normalised distance $z/L$, as in Figure \ref{fig:ratio-step} but for the case of a pulsed pump source. The transient (superradiance) intensity grows at a significantly higher rate than the steady-state unsaturated maser until $z\simeq 0.68L$ where the ratio reaches a peak of $I_{\mathrm{trans}}/I_{\mathrm{maser}}\approx 10^{10}$ (note the logarithmic scale on the vertical axis). This is in spite of the exponential growth of the unsaturated maser as a function of $z$ (see equation (\ref{eq:IntensityMaserWeak})). The increase of only a factor of $\sim3$ in the inversion level (see Figure \ref{fig:pulse}) was sufficient to bring an ``amplification'' of several orders of magnitude in intensity as superradiance sets in. A decline in the ratio is seen past its peak as the exponential gain of the maser catches up with that of superradiance (which scales as $\propto z^2$).}
    \label{fig:ratio-pulse}
\end{figure}

To facilitate the comparison between the superradiance and steady-state unsaturated maser intensities, we show their ratio (i.e., $I_{\mathrm{trans}}/I_{\mathrm{maser}}$) as a function of fractional distance $z/L$ in Figure \ref{fig:ratio-pulse}. As we move along the sample from $z/L\approx 0$, the transient (superradiance) intensity grows at a significantly higher rate than the unsaturated maser. This is in spite of the exponential growth of the latter as a function of $z$ (see equation (\ref{eq:IntensityMaserWeak})), which is an indication of the increased efficiency of a highly coherent process such as superradiance in releasing energy. As can be seen in the figure (note the logarithmic scale on the vertical axis), the increase of only a factor of $\sim3$ in the inversion level was sufficient to bring an ``amplification'' of several orders of magnitude in intensity as superradiance sets in. This is because in the process the column density of the inverted population exceeded the critical level. As a result, the ratio peaks near $z\simeq 0.68L$ at $I_{\mathrm{trans}}/I_{\mathrm{maser}}\approx 10^{10}$, where the exponential gain of the maser catches up with that of superradiance (which scales as $\propto z^2$). This continuing trend is also responsible for the decline afterwards at larger $z/L$ values. 

\begin{figure}
    \centering
    \includegraphics[width=\columnwidth]{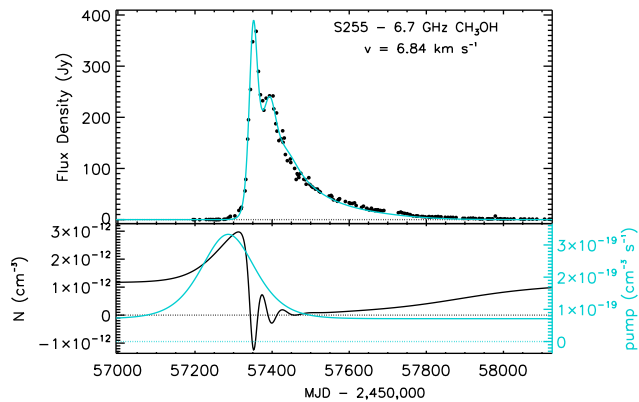}
    \caption{Superradiance model for the S255IR-NIRS3 6.7~GHz methanol flare at $v_{\mathrm{lsr}} = 6.84~\mathrm{km~s}^{-1}$ \citep{Szymczak2018b,Rajabi2019}. In the top panel of the figure the light/cyan solid curve is for the model fit to the data (dots), while in the bottom panel the temporal evolution for the pump (light/cyan curve, vertical axis on the right) and the population inversion density (dark/black curve, vertical axis on the left) are shown. The parameters of the sample are $L=2.17\times 10^{15}$~cm (imposing a Fresnel number of unity), $T_1=1.64 \times 10^{7}$~s and $T_2 =1.55 \times 10^6$~s, while the inversion level prior to the appearance of the pump pulse corresponds to approximately $0.1~\mathrm{cm}^{-3}$ for a molecular population spanning a velocity range of $1~\mathrm{km~s}^{-1}$. The duration of the flare ($\approx 300$~days) allows us to determine a peak column density for the inverted population of $nL\simeq 6.5\times 10^3\,\mathrm{cm}^{-2}$. The SR flux density is scaled to the data.}
    \label{fig:S255-fit}
\end{figure}

This behaviour is well suited to explain observations that show very intense and fast radiation flares appearing in maser-hosting regions. Such an example is presented in Figure \ref{fig:S255-fit}, where the light curve of a 6.7 GHz methanol flare observed in S255IR-NIRS3 is shown for one velocity channel ($v=6.84\,\mathrm{km}\,\mathrm{s}^{-1}$, the channel width is $0.44\,\mathrm{km}\,\mathrm{s}^{-1}$; see \citealt{Szymczak2018b} and \citealt{Rajabi2019} for more details). This source is well known for its maser emission and strong variability. In the top panel of the figure the light/cyan solid curve is for the model fit to the data (dots), while in the bottom panel the temporal evolution for the pump (light/cyan curve, vertical axis on the right) and the population inversion density (dark/black curve, vertical axis on the left) are shown. The parameters for the superradiance model are the same as those used in Figure \ref{fig:pulse}, except for a slight increase in the length of the sample ($L=2.17\times 10^{15}$~cm) and decrease in the constant pump level ($\Lambda_0=7.1\times 10^{-20}\,\mathrm{cm}^{-3}\,\mathrm{s}^{-1}$). The duration of the flare ($\approx 300$~days) allows us to determine a peak column density for the inverted population of $nL\simeq 6.5\times 10^3\,\mathrm{cm}^{-2}$. This event, which started at approximately MJD~57300 (2015 August), has been closely monitored by \citet{Szymczak2018b} and was linked to an infrared (IR) burst that started on MJD~57188 (2015 mid-June, \citealt{Caratti2017}). We note the time delay between the pump pulse and the superradiance transient in our model. 

This flare is consistent with the scenario discussed above, where a region hosting a weak unsaturated maser suddenly erupts into a very powerful superradiance transient event before returning to its initial state. The aforementioned IR burst would presumably be responsible for the pump and population inversion increases. \citet{Rajabi2019} discuss this superradiance model further, as well as those for two other sources.

We note that \citet{Uchiyama2020} have recently published a $Ks$ band IR light curve obtained for S255IR-NIRS3 at the epoch of the flare presented in Figure \ref{fig:S255-fit}. Their monitoring show a slow-rising exponential-looking IR pulse increasing by a factor of approximately 20 (3.4 mag; see their Fig. 3). Although their observations were obtained at a shorter wavelength, it is reasonable to assume that it shares a common origin with the pulse responsible for maser pumping. As noted in \citet{Rajabi2019}, the results of our superradiance models are fairly insensitive to the details of the pump pulse. Most notably, its shape is not transmitted to the intensity flare, as seen in Figure \ref{fig:S255-fit}. We have verified this and the quality of the fit using a slow-rising exponential pump pulse approximately mimicking the IR observations of \citet{Uchiyama2020}. All parameters in the model were kept unchanged except for the amplitude of the pump, which increased by a factor of $\sim6$ to $7$ instead of $~\sim3$ for the fit presented in Figure \ref{fig:S255-fit}. It is rewarding to note that the observations of \citet{Uchiyama2020} in the $Ks$ band show a likely increase in the pump pulse amplitude that is greater than needed for our model.

\section{Summary}\label{sec:conclusion}

We are now in a position to summarize the different aspects that characterise superradiance and its relationship to the maser action. The one overarching point around our analysis is that both phenomena can be shown to emerge from the same theoretical framework, as we did in Section \ref{sec:theory} following the literature published in the quantum optics community during the development of superradiance theory  \citep{Arecchi1970,MacGillivray1976,Feld1980,Gross1982,Benedict1996}. Although the presence of neither the stimulated emission process, which is at the root of the maser action \citep{Elitzur1992,Gray2012}, nor the entangled Dicke states, used by \citet{Dicke1954} to introduce superradiance, are readily apparent in the Maxwell-Bloch equations (i.e., equations (\ref{eq:dN/dt})-(\ref{eq:dE/dt})), their solution in different regimes allowed us to clearly identify when the two phenomena could be used to describe the ensuing radiation intensity. That is, we found in the quasi-steady state limit the signature of the maser action, while the manifestation of superradiance was a characteristic of the transient regime. 

More precisely, we can surmise the following points:
\begin{itemize}
    \item The masing action and superradiance are not competing phenomena in the radiation process but rather, as the previous comments imply, are complementary and define two distinct limits characterizing the intensity of radiation (i.e., the quasi-steady and transient limits, respectively).
    \item Coherent interactions which are at the root of superradiance can be established when the characteristic time-scales $T_\mathrm{R}$ and $\tau_\mathrm{D}$ are shorter than those pertaining to the non-coherent relaxation and dephasing processes. This sets a critical threshold for the inverted column density, above which the system acts as a single entangled quantum mechanical system and superradiance ensues. These conditions are met during the fast transient limit of the Maxwell-Bloch equations. 
    \item On the other hand, in the quasi-steady state limit of the Maxwell-Bloch equations the time-scales for temporal variations in the polarisation amplitude and inverted population density are very long compared to those of the non-coherent mechanisms. As a result, relaxation and dephasing processes randomly perturb the individual oscillating molecular dipoles and inhibit the establishment of a macroscopic dipole. This leads to low levels of polarisation and coherence in the system. The ensuing radiation exhibits characteristics of an unsaturated maser (i.e., an exponential growth with $z$) or a saturated maser (i.e., an amplification linear in $z$).
    \item The steady-state unsaturated maser regime is realised when the intensity is below the saturation intensity $I_{\mathrm{sat}}$, whereas a saturated maser is established whenever the intensity $I\gtrsim I_{\mathrm{sat}}$. Furthermore, the transition from the unsaturated maser to the saturated maser modes was also found to happen at a critical length $z_{\mathrm{crit}}$ where, not only the radiation intensity becomes approximately equal to $I_{\mathrm{sat}}$ but, the inverted column density reaches the aforementioned critical threshold. This coincides with the onset of superradiance in the system. The transient regime associated with superradiance can thus be seen as the intermediary between the two steady-state maser domains.
    \item Superradiance is characterised by oscillations in the polarisation amplitude and inverted population density (including negative values), a significant increase in output intensity that scales with $z^2$ (or $N^2$ in an homogeneous medium), as well as the scaling of both the characteristic time-scale $T_\mathrm{R}$ and delay time $\tau_\mathrm{D}$ with $N^{-1}$.
    \item The strong levels of coherence established during the superradiance regime makes it a highly efficient process for damping away the energy stored in the (inverted) system. This also explains the gain in radiation intensity attained during the superradiance phase relative to the two maser regimes, and shows that it is well suited to explain observations that reveal intense and fast radiation flares in maser-hosting regions.
\end{itemize}

Our analysis established the need to consider both the masing action and superradiance when studying, or trying to explain, the characteristics of the radiation emanating from a gas hosting a population inversion. This is because the behaviour of such a system is markedly different when undergoing the distinct phases defined by these mechanisms. This should not be surprising as both stimulated emission, which is at the root of the masing action, and superradiance equally are fundamental physical processes pertaining to the interaction between matter and radiation \citep{Feld1980}. Superradiance, being a generalisation of the spontaneous emission phenomenon to an entangled group of atoms/molecules, leads to significantly increased emission rates and is thus bound to play an important role in characterising the radiation intensity whenever conditions allow for quantum mechanical coherence to develop \citep{Dicke1954,Dicke1964}.

\section*{Acknowledgements}
M.H.'s research is funded through the Natural Sciences and Engineering Research Council of Canada Discovery Grant RGPIN-2016-04460. M.H. is grateful for the hospitality of Perimeter Institute where part of this work was carried out. F.R.'s research at Perimeter Institute is supported in part by the Government of Canada through the Department of Innovation, Science and Economic Development Canada and by the Province of Ontario through the Ministry of Economic Development, Job Creation and Trade. F.R. is in part financially supported by the Institute for Quantum Computing.

\bibliographystyle{mnras}
\bibliography{SR-bib} 

\appendix
\section{Strong dephasing}\label{sec:dephasing}

In Sections \ref{subsec:quasisteady} and \ref{subsec:transient}, we discussed two different limits of the Maxwell-Bloch equations, where temporal variations in the polarisation amplitude and population inversion density relative to $T_1$ and $T_2$ were both either very slow or very fast. In actual systems, polarisation dephasing processes often pose more restriction on the evolution of coherence compared to the non-coherent relaxation mechanisms. More precisely, it is normally the case that $T_2 \ll T_1$ (for example, elastic collisions will affect $P^+$ but not $n^\prime$) and consequently we can have situations when 
\begin{equation}
\frac{\partial n^\prime}{\partial\tau} \gg \frac{n^\prime}{T_1} \quad\mathrm{and} \quad \frac{\partial P^+}{\partial\tau} \ll \frac{P^+}{T_2}. \label{eq:ASElimit}    
\end{equation}
When these conditions hold in a system equations (\ref{eq:dN/dt})-(\ref{eq:dE/dt}), at resonance, can be simplified to
\begin{align}
& \frac{\partial n^\prime}{\partial\tau} = \frac{i}{\hbar}\left(P^+E^+-E^-P^-\right)\label{eq:dN/dt-ASE} \\   
& \frac{P^+}{T_2} = \frac{2id^2}{\hbar}E^-n^\prime  \label{eq:dP/dt-ASE} \\
& \frac{\partial E^+}{\partial z} = \frac{i\omega_0}{2\epsilon_0c} P^-, \label{eq:dE/dt-ASE}
\end{align}
for which the time-derivative of the polarisation amplitude was neglected relative to $P^+ /T_2$. To simplify matters further the pump terms for the polarisation and inversion density were also set to zero, while the contribution of $n^\prime/T_1$ was further neglected in equation (\ref{eq:dN/dt-ASE}). Inserting equation (\ref{eq:dP/dt-ASE}) in equations (\ref{eq:dN/dt-ASE}) and (\ref{eq:dE/dt-ASE}) we get 
\begin{align}
& \frac{\partial n^\prime}{\partial\tau} = -\frac{4n^\prime d^2T_2}{\hbar^2}E^-E^+\label{eq:dN/dt-ASE2} \\
& \frac{\partial E^+}{\partial z} = \frac{\omega_0d^2T_2}{\epsilon_0c\hbar} E^+n^\prime,\label{eq:dE/dt-ASE2} 
\end{align}
which upon applying the definition of the radiation intensity $I$ transform to  
\begin{align}
& \frac{\partial n}{\partial\tau} = -\frac{8nd^2T_2}{c\epsilon_0\hbar^2}I \label{eq:dN/dt-ASE3} \\   
& \frac{\partial I}{\partial z} = \alpha I. \label{eq:dE/dt-ASE3}
\end{align}

Equation (\ref{eq:dE/dt-ASE3}) is identical to the relation obtained for the spatial variations in intensity pertaining to the quasi-steady state limit, with the exception that, here, the gain coefficient $\alpha$ is a function of the retarded time through its dependency on $n$ (see equation (\ref{eq:alpha})). We therefore see that the conditions given in equations (\ref{eq:ASElimit}) lead to a non-coherent amplification process consistent with the masing action, not superradiance \citep{Benedict1996}.

\section{Lethargic gain  and critical threshold}\label{sec:lethargic}

We consider a situation where the intensity of radiation and the polarisation are weak, as will apply whenever $\tau\rightarrow0$ and/or $z$ is not in the region hosting a saturated maser. In such cases we can safely assume that $\partial n^\prime/\partial \tau\approx 0$, while the product $P^\pm E^\pm$, being of second order in size, can be neglected in the first of Maxwell-Bloch equations (equation (\ref{eq:dN/dt-Ideal})) with the result that $n^\prime=n^\prime_0=\Lambda_0T_1$. As was mentioned in Section \ref{subsec:step}, we cannot set $\partial P^+/\partial \tau\approx 0$ in equation (\ref{eq:dP/dt-ideal}) since quantum perturbations can lead to substantial changes in polarisation when it is small to start with. Taking these considerations into account and performing a spatial derivative on equation (\ref{eq:dP/dt-ideal}), the Maxwell-Bloch equations are transformed to    
\begin{align}
    & \frac{\partial^2 P^+\left(z,\tau\right)}{\partial z\partial\tau} = \frac{1}{LT_\mathrm{R}}P^+\left(z,\tau\right)-\frac{1}{T_2}\frac{\partial P^+\left(z,\tau\right)}{\partial z}\label{eq:d2P/dzdt} \\
    & \frac{\partial E^+\left(z,\tau\right)}{\partial z} = \frac{i\omega_0}{2\epsilon_0 c}P^-\left(z,\tau\right),\label{eq:dE/dz-leth}
\end{align}
where equation (\ref{eq:TR}) was used with $n=2n^\prime$. These equations can be solved using Laplace transforms, with the $\tau\leftrightarrow s$ and $z\leftrightarrow u$ correspondences, and the initial conditions
\begin{align}
    & P^+\left(z=0,\tau\right) = P^+\left(z,\tau=0\right) = \frac{n_0 d\,\theta_0}{2} \label{eq:P_initial} \\
    & E^+\left(z=0,\tau\right) = 0\label{eq:E_initial}
\end{align}
used for the numerical examples of Section \ref{subsec:step}. 

First transforming equation (\ref{eq:d2P/dzdt}) for $P^-\left(u,s\right)$, followed by its insertion into the equation for $E^+\left(u,s\right)$ (from equation (\ref{eq:dE/dz-leth})) we find
\begin{equation}
    E^+\left(u,s\right) = \frac{i \hbar\theta_0}{2dLT_\mathrm{R}}\frac{\left(s+1/T_2\right)}{u^2s\left(s+1/T_2-1/uLT_\mathrm{R}\right)}.\label{eq:E+solution}
\end{equation}
Performing a first inverse Laplace transform to recover the retarded time parameter yields
\begin{equation}
    E^+\left(u,\tau\right) = \frac{i\hbar\theta_0}{2dT_2}\frac{\alpha}{2}\left[\frac{1}{u\left(u-\alpha/2\right)}-\frac{\alpha}{2}e^{-\tau/T_2}\frac{e^{\tau/uLT_\mathrm{R}}}{u^2\left(u-\alpha/2\right)}\right],\label{eq:E+solution2}
\end{equation}
where we used equation (\ref{eq:alpha-TR}) for $\alpha$. For $\tau>T_2$, as is the case in the steady-state regime, the last term in equation (\ref{eq:E+solution2}) becomes negligible and the final inverse Laplace transform gives for a solution
\begin{equation}
    E^+\left(z,\tau\right) = \frac{i\hbar\theta_0}{2dT_2}\left(e^{\alpha z/2}-1\right).\label{eq:E+solution3}
\end{equation}
For $z\ll L$ we find that $E^+\left(z,\tau\right)\propto z$, which corresponds to a coherent behaviour (i.e., the intensity scales with $z^2$), while for larger $z$ values equation (\ref{eq:E+solution3}) falls into the unsaturated maser regime (see equation (\ref{eq:IntensityMaserWeak})). 

We also note that when $T_2\gg T_\mathrm{R}$ equation (\ref{eq:E+solution2}) can be approximated to 
\begin{equation}
    E^+\left(u,\tau\right) \simeq \frac{i\hbar\theta_0}{2LdT_\mathrm{R}}e^{-\tau/T_2}\frac{e^{\tau/uLT_\mathrm{R}}}{u^2},\label{eq:E+solution4}
\end{equation}
which has for solution
\begin{equation}
     E^+\left(z,\tau\right) \simeq \frac{i\hbar\theta_0}{2dT_\mathrm{R}}e^{-\tau/T_2}\sqrt{\frac{zT_\mathrm{R}}{L\tau}}I_1\left(2\sqrt{\frac{z\tau}{LT_\mathrm{R}}}\right),\label{eq:E+solution5}
\end{equation}
where we used
\begin{equation}
    \mathcal{L}^{-1}\left\{\frac{e^{a/u}}{u^n}\right\} = \left(\frac{z}{a}\right)^\frac{n-1}{2}I_{n-1}\left(2\sqrt{az}\right),\label{eq:BesselI}
\end{equation}
with $I_n\left(x\right)$ the modified Bessel function of the first kind of order $n$ \citep{Gradshteyn1980}. 

For $z\tau\gg L T_\mathrm{R}$ we can use the asymptotic expansion $I_1\left(x\right)=e^x/\sqrt{2\pi x}$ as $x\rightarrow\infty$ to obtain \citep{Gradshteyn1980}
\begin{equation}
     E^+\left(z,\tau\right) \approx \frac{i\hbar\theta_0}{2dT_\mathrm{R}}e^{-\tau/T_2}\left(\frac{z}{L}\right)^{1/4}\left(\frac{T_\mathrm{R}}{\tau}\right)^{3/4}\frac{e^{2\sqrt{z\tau/LT_\mathrm{R}}}}{\sqrt{4\pi}}.\label{eq:E+_lethargic}
\end{equation}
This particular form for the growth of the electric field (i.e., with the exponent proportional to $\sqrt{z}$) is different from the more common Beer's Law (i.e., with the exponent proportional to $z$, as for the unsaturated maser) and is commonly called ``lethargic gain.'' For the problem at hand, this regime is restricted to $\tau\lesssim T_2$ \citep{Benedict1996}. 

We note that the extension of equation (\ref{eq:E+solution3}) to the unsaturated maser regime allows us to write (see equation (\ref{eq:IntensityMaserWeak}))
\begin{equation}
    I_0 = \frac{c\epsilon_0}{2}\left(\frac{\hbar\theta_0}{2dT_2}\right)^2.\label{eq:I0}
\end{equation}
We are then in a position to find a relation for the critical level for the column density of the population inversion by equating the unsaturated maser intensity with the saturation intensity $I_\mathrm{sat}$ (equation (\ref{eq:Is})). Doing so yields
\begin{equation}
    \left(n_0L\right)_\mathrm{crit} = \frac{4\pi}{3\lambda^2}\frac{\tau_\mathrm{sp}}{T_2}\ln{\left(\frac{T_2}{T_1\theta_0^2}\right)},\label{eq:nLcrit}
\end{equation}
which can be readily converted to either a critical length $z_\mathrm{crit}$ or a critical inverted density $n_{0,\mathrm{crit}}$ depending on whether the density or the length of the system are given, respectively.

\bsp	
\label{lastpage}
\end{document}